\documentclass[prd,twocolumn,superscriptaddress,showpacs,nofootinbib, amsmath, amssymb]{revtex4-1}

\usepackage{graphicx}
\usepackage{hyperref}
\usepackage{color}
\usepackage[utf8]{inputenc}

\begin{document}

\title{Modeling evolution of dark matter substructure and annihilation
boost}

\author{Nagisa Hiroshima}
\affiliation{Institute for Cosmic Ray Research, University of Tokyo,
Kashiwa, Chiba 277-8582, Japan}
\affiliation{Institute of Particle and Nuclear Studies, High Energy
Accelerator Research Organization (KEK), Tsukuba, Ibaraki 305-0801,
Japan}
\author{Shin'ichiro Ando}
\affiliation{GRAPPA Institute, Institute of Physics, University of Amsterdam, 1098 XH Amsterdam, The Netherlands}
\affiliation{Kavli Institute for the Physics and Mathematics of the
Universe (Kavli IPMU, WPI), Todai Institutes for Advanced Study,
University of Tokyo, Kashiwa, Chiba 277-8583, Japan}
\author{Tomoaki Ishiyama}
\affiliation{Institute of Management and Information Technologies, Chiba
University, Chiba 263-8522, Japan}

\begin{abstract}
We study evolution of dark matter substructures, especially how they
 lose the mass and change density profile after they fall in
 gravitational potential of larger host halos.
We develop an analytical prescription that models the subhalo mass
 evolution and calibrate it to results of $N$-body numerical simulations
 of various scales from very small (Earth size) to large (galaxies to
 clusters) halos.
We then combine the results with halo accretion histories, and calculate
 the subhalo mass function that is physically motivated down to
 Earth-mass scales. 
Our results --- valid for arbitrary host masses and redshifts --- show
 reasonable agreement with those of numerical simulations at resolved
 scales.
Our analytical model also enables self-consistent calculations of the
 boost factor of dark matter annhilation, which we find to increase from
 tens of percent at the smallest (Earth) and intermediate (dwarfs)
 masses to a factor of several at galaxy size, and to become as large as
 a factor of $\sim$10 for the largest halos (clusters) at small
 redshifts.
Our analytical approach can accommodate substructures in the subhalos
 (sub-subhalos) in a consistent framework, which we find to give up to a
 factor of a few enhancement to the annihilation boost.
Presence of the subhalos enhances the intensity of the isotropic
 gamma-ray background by a factor of a few, and as the result, the
 measurement by Fermi Large Area Telescope excludes the annihilation
 cross section greater than $\sim$$4\times 10^{-26}$~cm$^3$~s$^{-1}$ for
 dark matter masses up to $\sim$200~GeV.
\end{abstract}

\date{March 28, 2018; revised \today}

\maketitle

\section{Introduction}
\label{sec:intro}

There is strong evidence for the existence of dark matter, such as the
distribution of matter in the
Universe~\cite{Peebles:1982ff,Ade:2015xua}, rotation curves of
galaxies~\cite{vanAlbada:1984js,Salucci:2002jg}, bullet
clusters~\cite{Clowe:2003tk}, etc. 
In spite of the efforts to unveil the nature of the dark matter, however, our
knowledge about it is still limited.
Many models of particle dark matter have been proposed, and among them,
weakly interacting massive particles (WIMPs) are one of the best studied
in accordance with supersymmetric extensions of the standard
model~\cite{Jungman:1995df}.
If dark matter is made of new particles such as WIMPs, which have small
but finite interaction with standard model sector, we expect them to be 
detected through the observations of gamma rays from self-annihilation
of dark matter particles~\cite{Gaskins:2016cha}.

Dark matter forms virialized objects --- dark matter halos, which give
some hints about its nature.
For example, they encode information of scattering between dark matter
particles and the standard model particles in the early Universe, through the minimum
halo mass being predicted to be $10^{-12}$--$10^{-3} M_\odot$ for the
supersymmetric neutralino~\cite{Hofmann:2001bi, Green:2003un,
Profumo:2006bv, Diamanti:2015kma}.
Halos grow larger and larger by merging to each other and accreting
smaller ones, leaving imprints of dark matter properties in their
hierarchical structures.
Smaller halos that are accreted onto larger (host) halos are referred to
as subhalos or substructures.
Once subhalos are trapped by their hosts, they lose their mass through
gravitational tidal force while orbiting.
With given properties of the host and subhalos at their accretion,
we can determine the tidal mass loss of the subhalos and remaining
structures after some orbiting time.
This procedure is studied through the
analytical~\cite{vandenBosch:2004zs, Giocoli:2007gf, Jiang:2014nsa},
semi-analytical~\cite{Penarrubia:2004et} and numerical~\cite{Gao:2004au,
Diemand:2007qr, Giocoli:2007uv, Dolag:2008ar, Springel:2008cc,
vandenBosch:2017ynq} approaches.

Subhalos remaining in their host are {\it boosters} for indirect
detection experiments of particle dark
matter~\cite{Diemand:2006ik, Strigari:2006rd, Pieri:2007ir,
Jeltema:2008ax, Ishiyama:2014uoa, Bartels:2015uba}, especially for
gamma-ray telescopes such as the Fermi Large Area Telescope (LAT).
In order to discuss the evolution of subhalo abundance, mass
distribution, and density profile, and to estimate the substructure
boost, analytical modeling is a powerful tool since they do not suffer
from resolution limits.
We can cover wide range of magnitude in both the host-halo mass and
the mass ratio of the hosts to subhalos in the analytical calculations.

In this paper, we discuss properties of subhalos after tidal stripping,
and as one of the applications, the boost factor for the gamma-ray
signals from dark matter annihilation.
This study updates calculations of the substructure boost by
Reference.~\cite{Bartels:2015uba} in various aspects.
In order to access the properties of the subhalos after accretion, we
follow an analytical approach in Reference.~\cite{Jiang:2014nsa}, which
considers the mass loss of the subhalos due to the tidal stripping under
the potential of the host halos.
This analytical model is physically motivated although it has simplified
some aspects of tidal stripping.
We include the host mass and redshift dependence of the tidal stripping
for the purpose of improving the accuracy of the models. 
Then, we consistently take evolutions of the host and subhalos into
account in calculations of their properties. 
After modeling of the tidal mass loss of subhalos, we calculate the
boost factors of the subhalos for the gamma-ray signals from dark matter
annihilation as well as mass function of subhalos.

The structure of this article is as follows. 
In Sec.~\ref{sec:subhalostructure}, we explain the ways to derive the
properties of subhalos after tidal stripping from quantities at
the accretion time. 
In Sec.~\ref{sec:stripping}, we derive the host mass and redshift
dependence of the subhalo mass-loss rate.
In Sec.~\ref{sec:result}, we show applications to the observational
signatures such as subhalo mass function and annihilation boost factor. 
We then discuss implications for the isotropic gamma-ray background in
Sec.~\ref{sec:discussion}, and summarize our findings in
Sec.~\ref{sec:conclusion}.
Throughout the paper, we adopt cosmological parameters from
Reference.~\cite{Ade:2015xua} (Table 4, ``TT+lowP+lensing''), and use
``$\ln$'' and ``$\log$'' to represent natural and 10-base logarithmic
functions, respectively.

\section{Density profile of subhalos}
\label{sec:subhalostructure}

Dark matter halos have evolved by merging and accretion.
After accretion onto their hosts, subhalos lose their mass due to tidal
stripping while they are orbiting in their host's gravitational
potential.
In this section, we show that the properties of subhalos after tidal
stripping can be determined given the mass $m_\mathrm{acc}$ at accretion
redshift $z_\mathrm{acc}$ for given host halos, on a statistical basis.
Starting from ($m_\mathrm{acc},z_\mathrm{acc}$), we can calculate the
subhalo mass at a redshift $z_0$, denoted as $m_0$, by integrating its
mass-loss rate $\dot{m}$ from accretion redshift $z_\mathrm{acc}$ to
$z_0$.
We parameterize the mass-loss rate as
\begin{equation}
 \dot m(z) = - A\frac{m(z)}{\tau_{\rm
  dyn}(z)}\left[\frac{m(z)}{M(z)}\right]^{\zeta},
  \label{eq:mdot}
\end{equation}
where $\tau_{\rm dyn}(z)$ is the dynamical
timescale~\cite{Jiang:2014nsa}.
The evolution of the host mass $M(z)$ is discussed in
References.~\cite{Correa:2014xma}, and is also summarized in
Appendix~\ref{app:Mass evolution of host halos}.
Parameters $A$ and $\zeta$ are taken to be constants in
Reference.~\cite{Jiang:2014nsa}, but in a more realistic case, both of them
should depend on the host mass $M(z)$ and the redshift $z$.
We derive the dependence following the analytical discussion in
Reference.~\cite{Jiang:2014nsa} with several updates in the next section.

In this section, we show how density profiles of the subhalos including
a scale radius $r_s$ and a characteristic density $\rho_s$ evolve,
associated to the evolution of the subhalo mass from $m_\mathrm{acc}$ at
$z_\mathrm{acc}$ to $m_0$ at $z_0$.
Throughout our calculations, we adopt the Navarro-Frenk-White
(NFW) density profile~\cite{Navarro:1996gj} up to a truncation radius
$r_t$, and zero beyond:
\begin{equation}
 \rho(r) = 
  \left\{
   \begin{array}{lll}
    \rho_s r_s^3 / [r(r+r_s)^2], & \mbox{for} & r\le r_t, \\
    0, & \mbox{for} & r>r_t. \\
   \end{array}
  \right.
\end{equation}

First, we determine $\rho_s$ and $r_s$ at the accretion redshift
$z_\mathrm{acc}$.
As it was a {\it field} halo (i.e., a halo that is not in a larger
halo's gravitational potential) when accreted, we first determine the
virial radius $r_\mathrm{vir,acc}$ at $z_\mathrm{acc}$ from the mass of
the subhalo at accretion $m_\mathrm{acc}$:
\begin{equation}
m_\mathrm{acc} = \frac{4\pi}{3} \Delta_c(z_a)\rho_c(z_\mathrm{acc})
 r_\mathrm{vir,acc}^3,
\end{equation}
where $\Delta_c = 18\pi^2+82d-39d^2$, $d = \Omega_m(1+z_\mathrm{acc})^3
/ [\Omega_m(1+z_\mathrm{acc})^3+\Omega_\Lambda]-1$~\cite{Bryan:1997dn},
and $\rho_c(z_\mathrm{acc})$ is the critical density at
$z_\mathrm{acc}$. 
The scale radius is determined by $r_{s,\mathrm{acc}} =
r_{\mathrm{vir,acc}}/c_\mathrm{vir,acc}$ at $z_\mathrm{acc}$ once a
concentration parameter $c_\mathrm{vir,acc}$ is given. 
The concentration follows the log-normal distribution, whose mean is
obtained in, e.g., Reference.~\cite{Correa:2015dva}, which is summarized in
Appendix~\ref{app:concentration}.
Note that Reference.~\cite{Correa:2015dva} defines the concentration as a
function of halo masses measured in $M_{200}$, defined as enclosed mass
in a radius within which the average density is 200 times the critical
density.
The virial concentration parameter $c_\mathrm{vir,acc}$ is obtained by a
conversion between different definitions of mass~\cite{Hu:2002we},
followed by $c_\mathrm{vir,acc} = c_{200,\mathrm{acc}}
{r_\mathrm{vir,acc}} / {r_\mathrm{200,acc}}.$
For the rms of the log-normal distribution, we adopt $\sigma_{\log c}
=0.13$~\cite{Ishiyama:2011af}.
The characteristic density $\rho_{s,\mathrm{acc}}$ is then determined
from
\begin{equation}
\rho_{s,\mathrm{acc}} = \frac{m_\mathrm{acc}}{4\pi
 r_{s,\mathrm{acc}}^3f(c_\mathrm{vir,acc})},
\end{equation}
where 
\begin{equation}
\label{eq:fc}
f(c)  = \ln(1+c)-\frac{c}{1+c}.
\end{equation}

The set of parameters $(r_{s,\mathrm{acc}}, \rho_{s,\mathrm{acc}})$ is
related to the maximum circular velocity $V_\mathrm{max}$ and radius
$r_\mathrm{max}$ at which the circular velocity reaches maximum through
\begin{eqnarray}
r_s & = & \frac{r_\mathrm{max}}{2.163},
\label{defrmax}\\
\rho_s &= &\frac{4.625}{4\pi
 G}\left(\frac{V_\mathrm{max}}{r_s}\right)^2.
\label{defVmax}
\end{eqnarray}
Reference~\cite{Penarrubia:2010jk} derived the relation between the
subhalo properties before and after the tidal stripping by following
the evolution of $V_\mathrm{max}$ and $r_\mathrm{max}$. 
The relation between the ($V_\mathrm{max}$, $r_\mathrm{max}$) at
accretion redshift $z_\mathrm{acc}$ and those at the arbitrarily chosen
observation redshift $z_0$, in terms of the mass ratio after and
before tidal stripping $m_0/m_\mathrm{acc}$, is
\begin{eqnarray}
\frac{V_\mathrm{max,0}}{V_\mathrm{max,acc}}&=&\frac{2^{0.4}(m_0/m_\mathrm{acc})^{0.3}}{\left(1+m_0/m_\mathrm{acc}\right)^{0.4}},\\
\frac{r_\mathrm{max,0}}{r_\mathrm{max,acc}}&=&\frac{2^{-0.3}(m_0/m_\mathrm{acc})^{0.4}}{\left(1+m_0/m_\mathrm{acc}\right)^{-0.3}},
\end{eqnarray}
for inner density profile proportional to $r^{-1}$, as is the case of
the NFW.
Then, we can determine $r_{s,0}$ and $\rho_{s,0}$ at $z=z_0$ through
$V_\mathrm{max}$ and $r_\mathrm{max}$ in Eqs.~(\ref{defrmax}) and
(\ref{defVmax}).
Finally, the truncation radius $r_{t,0}$ is determined from $m_0$,
$\rho_{s,0}$ and $r_{s,0}$ by solving 
\begin{equation}
m_0 = 4\pi \rho_{s,0}r_{s,0}^3
f\left(\frac{r_{t,0}}{r_{s,0}}\right).
\end{equation}
We remove the subhalos with $r_{t,0} / r_{s,0} < 0.77$ from further
consideration, as it is usually assumed that the subhalos satisfying
this condition are completely disrupted~\cite{Hayashi:2002qv}. 
(But see Reference.~\cite{vandenBosch:2017ynq} for a claim otherwise.)

To summarize, following the prescription in this section (and the
mass-loss rate $\dot m$ discussed in the next section), we can determine
the density profile of the subhalos after tidal stripping at an
arbitrary redshift $z_0$ up to scatter of the concentration-mass
relation, given the mass and redshift of accretion, $m_{\rm acc}$ and
$z_{\rm acc}$.
Combined with distribution of $m_{\rm acc}$ and $z_{\rm acc}$ that is
obtained with the extended Press-Schechter formalism~\cite{Yang:2011rf}
(summarized in Appendix~\ref{ssec:Subhalo accretion rate}), we can
compute the statistical average of subhalo quantities of various
interests.
Among them, we discuss the subhalo mass functions and annihilation boost
factor in Sec.~\ref{sec:result}.

\section{Tidal stripping}
\label{sec:stripping}

The subhalo mass-loss rate $\dot{m}$, as can be seen in Eq.~(\ref{eq:mdot}), should depend on both the
redshift $z$ and the host mass $M(z)$, since the subhalo evolution is
determined by the tidal force of their host.
Following Reference.~\cite{Jiang:2014nsa}, by assuming that tidal stripping of
the subhalos occur in one complete orbital period and there are no lags
between the subhalo accretion and the tidal stripping of those accreted,
we estimate the mass-loss rate of the accreted subhalos on a certain
host at any redshift in an analytical way.
We also show consistency of our results with those obtained by numerical
simulations.

\subsection{Analytical model}
\label{AM}

The mass loss $\dot{m}(z)$ of any subhalo is approximated as
\begin{equation}
\dot{m}=\frac{m-m(r_t)}{T_r},
\label{dotmdef}
\end{equation}
where $T_r$, $m$, and $m(r_t)$ are the orbital period, the virial mass
of the subhalo just after accretion, and the mass enclosed in the tidal
truncation radius $r_t$ of the subhalo, respectively.
In order to determine the orbit of the subhalo, we draw the orbit
circularity $\eta$ at infall, and radius of the circular orbit $R_c$
from distribution functions for each parameter:
\begin{equation}
P(R_c)=\begin{cases} 5/2 &(0.6\leq{R_c}/R_\mathrm{vir}\leq1.0),\\
0  &(\mathrm{otherwise}),
\end{cases}
\label{Rcdist}
\end{equation}
\begin{equation}
 P({\eta})=C_0(M,z)\eta^{1.05}(1-\eta)^{C_1(M,z)},
\label{etadist}
\end{equation}
where
\begin{eqnarray}\label{c0eta}
C_0&=&3.38\left(1+0.567\left[\frac{M}{M_*(z)}\right]^{0.152}\right),\\ 
C_1&=&0.242\left(1+2.36\left[\frac{M}{M_*(z)}\right]^{0.107}\right),\label{c1eta}
\end{eqnarray}
\begin{equation}
\log\left[\frac{M_*(z)}{h^{-1}M_\odot}\right]=12.42-1.56z+0.038z^2.
\label{eq:Mstar}
\end{equation}
We note that Eqs.~(\ref{etadist})--(\ref{eq:Mstar}) are calibrated with
simulations up to $z = 7$~\cite{Wetzel:2010kz}.
Pairs of $\eta$ and $R_c$ correspond to the pairs of the angular
momentum $L$ and the total energy $E$ of the orbiting subhalo as follows:
\begin{eqnarray}
E&=&\frac{1}{2}V_c^2+\Phi(R_c),\\
L&=& \eta R_cV_c,
\end{eqnarray}
where $V_c=(GM/R_c)^{1/2}$ is a velocity at the circular orbit.
The gravitational potential of the host $\Phi$ is 
\begin{equation}
\Phi(R)=-V_\mathrm{vir}^2\frac{\ln[1+c_{\rm vir}^{\rm host}
 R/R_\mathrm{vir}]}{f(c_{\rm vir}^{\rm host})R/R_\mathrm{vir}},
\end{equation}
with $V_{\rm vir} = (GM/R_{\rm vir})^{1/2}$ and $c_{\rm vir}^{\rm host}$
the host halo's virial velocity and virial concentration, respectively.
Here, we draw $c_{\rm vir}^{\rm host}$ from the log-normal distribution as
discussed in the previous section.

Next, we determine the orbital period, $T_r$, and the truncation radius
of the subhalo, $r_t$.
They are derived from the pericenter radius $R_p$ and the apocenter
radius $R_a$, which are obtained by solving
\begin{equation}
\frac{1}{R^2}+\frac{2[\Phi(R)-E]}{L^2}=0.
\end{equation}
The orbital period $T_r$ is then
\begin{equation}
T_r=2\int^{R_a}_{R_p}\frac{dR}{\sqrt{2[E-\Phi(R)]-L^2/R^2}}.
\end{equation}
The truncation radius $r_t$ is obtained by solving the equation
\begin{equation}
r_t=R_p\left[\frac{m(r_t)/M(<R_p)}{2+\frac{L^2}{R_pGM(<R_p)}-\left.\frac{d\ln{M}}{d\ln{R}}\right|_{R_p}}\right]^\frac{1}{3}.
\end{equation}
Assuming that $\rho_s$ and $r_s$ hardly change as the result of one
complete orbit after the infall, we specify the mass profile $m(r)$ up
to truncation radius $r_t$, and hence are able to compute the mass-loss
rate $\dot m$ with Eq.~(\ref{dotmdef}).

We made this simplified assumption of unchanged $\rho_s$ and $r_s$ in
order to capture the most relevant physics of tidal mass loss in our
analytical modeling.
According to Reference.~\cite{Penarrubia:2010jk}, however, $\rho_s$ and $r_s$
do change in one orbit by $\alt 50$\%.
Although we have neglected this effect in the model of tidal stripping, our
results show good agreements with those of N-body simulations as we show below.
This is likely due to the compensation of the changes of $\rho_s$ and
$r_s$ with those of $r_t$, and therefore, our simplification does not
affect our estimates about the tidal mass-loss of subahlos significantly.

\subsection{Numerical simulations}
\label{ssec:NS}

We have also calculated the tidal stripping of subhalos using $N$-body
simulations.
To cover a wide range of halo mass, we used five large cosmological
$N$-body simulations.  
Table~\ref{tab:simulations} summarizes the detail of these simulations.  
The $\nu^2$GC-S, $\nu^2$GC-H2~\cite{Ishiyama:2014gla}, and Phi-1 simulations cover halos with large mass ($\sim$$10^{11} M_{\odot}$). 
The Phi-2 simulation is for intermediate mass halos ($\sim$$10^{7}
M_{\odot}$).  
To analyze the smallest scale ($\sim$$10^{-6} M_{\odot}$), the
A\_N8192L800 simulation is used. 
The cosmological parameters of these simulations are $\Omega_m=0.31$,
$\lambda_0=0.69$, $h=0.68$, $n_s=0.96$, and $\sigma_8=0.83$, which are
consistent with an observation of the cosmic microwave background
obtained by the Planck satellite~\cite{Ade:2013zuv,Ade:2015xua} and
those adopted in the other sections of the present paper.
The matter power spectrum in the A\_N8192L800 simulation contained the
cutoff imposed by the free motion of dark matter particles with a mass
of 100~GeV~\cite{Green:2003un,Ishiyama:2014uoa}.  
Further details of these simulations are presented in
Reference.~\cite{Ishiyama:2014gla} and Ishiyama et al. (in preparation).

All simulations were conducted by a massively parallel TreePM code,
GreeM~\cite{Ishiyama:2009qn,Ishiyama:2012gs}.\footnote{\url{http://hpc.imit.chiba-u.jp/~ishiymtm/greem/}}
Halos and subhalos were identified by ROCKSTAR phase space halo and
subhalo finder~\cite{Behroozi:2011ju}. 
Merger trees are constructed by consistent tree
codes~\cite{Behroozi:2011js}.
The halo and subhalo catalogs and merger trees of the
$\nu^2$GC-S, $\nu^2$GC-H2, and Phi-1 simulations are publicly available at \url{http://hpc.imit.chiba-u.jp/~ishiymtm/db.html}.

\begin{table*}[h!t]
\caption{\label{tab:simulations}
Details of five cosmological $N$-body simulations used in this study.
 Here, $N$, $L$, and $m_{\rm p}$ are the total number of particles, box
 size, and mass of a simulation particle, respectively.}
\begin{ruledtabular}
\begin{tabular}{lccccc}
Name & $N$ & $L$ & Softening & $m_{\rm p}$ ($\rm M_{\odot}$) & Reference\\
\hline
$\nu^2$GC-S & $2048^3$ & 411.8 Mpc& 6.28 kpc & $3.2 \times 10^{8}$ & ~\cite{Ishiyama:2014gla,Makiya:2015spa}\\
$\nu^2$GC-H2 & $2048^3$ & 102.9 Mpc& 1.57 kpc & $5.1 \times 10^{6}$ & ~\cite{Ishiyama:2014gla,Makiya:2015spa}\\
Phi-1 & $2048^3$ & 47.1 Mpc& 706 pc & $4.8 \times 10^{5}$ & Ishiyama et al. (in prep)  \\
Phi-2 & $2048^3$ & 1.47 Mpc& 11 pc & 14.7 & Ishiyama et al. (in prep)  \\
A\_N8192L800 & $8192^3$ & 800.0 pc& $2.0 \times 10^{-4}$ pc & $3.7 \times 10^{-11}$ & Ishiyama et al. (in prep)  \\
\end{tabular}
\end{ruledtabular}
\end{table*}

\subsection{Comparison}

We calculate the mass-loss rate of the subhalos for various redshift
$z$ and the host mass $M_{\rm host}$ (defined as $M_{200}$).
First, we choose the subhalo mass at accretion $m_{\rm acc}$ uniformly in
a logarithmic scale between the smallest mass $10^{-6}M_\odot$ and the
maximum mass $0.1M(z_{\rm acc})$.
For each set of $m_{\rm acc}$ and $z_{\rm acc}$ (as well as $z$ and
$M_{\rm host}$), we calculate the mass-loss rate $\dot m$
following the prescription given in Sec.~\ref{AM}, by taking a Monte
Carlo appraoch; i.e., by drawing the concentration of the host halos,
subhalo concentration, circularity $\eta$, and radius of the circular
orbit $R_c$ of subhalos following the distributions of each of these
parameters.

\begin{figure}
 \begin{center}
  \includegraphics[width=10cm]{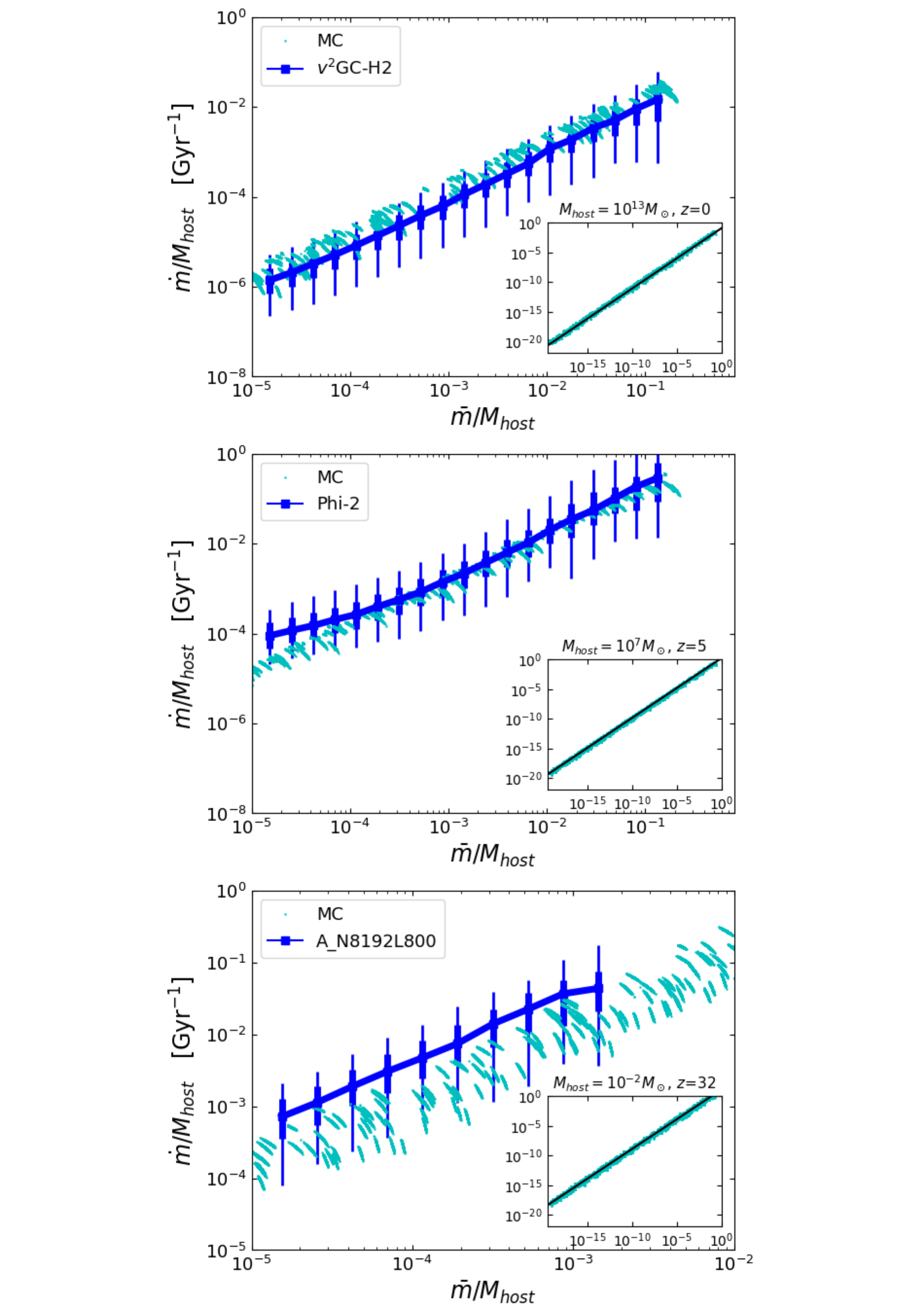}
\caption{Mass-loss rate of subhalos as a function of orbit-averaged
 subhalo mass $\overline m$ in units of the host mass $M_{\rm host}$ for
 $M_\mathrm{host}=10^{13}M_\odot$ and $z = 0$ (${\it top}$),
 $M_\mathrm{host}=10^7M_\odot$ and $z=5$ (${\it middle}$), and
 $M_\mathrm{host}= 10^{-2}M_\odot$ and $z=32$ (${\it bottom}$). Cyan points
 show the Monte Carlo simulation results. Blue squares with error bars
 show the results obtained by $N$-body simulations. Thick error bars
 correspond to the 50\% of the simulated halos around the median, while
  thin ones to the 90\%. We also show the results of the Monte Carlo
  simulations of wider mass range in inserted panels, which also include
  the fitting results with Eq.~(\ref{eq:mdot}), as overwritten solid lines
  on the Monte Carlo points.}
  \label{fig:mdotvsm}
 \end{center}
\end{figure}

In Figure.~\ref{fig:mdotvsm}, we show results of our Monte Carlo
simulations.
We find that for a large dynamic range of subhalo mass $m$ (over 19
orders of magnitude as shown in the insets) down to very small masses
such as $10^{-6} M_\odot$, a single power-law function
[Eq.~(\ref{eq:mdot})] gives a very good fit, which confirms the {\it
physical} origin of this relation, not just being a simple
phenomenological fit.

We compare the results of the Monte Carlo calculations to those
of the $N$-body simulations as described in Sec.~\ref{ssec:NS}, which
is also shown in Figure.~\ref{fig:mdotvsm} for $\overline{m}/M_{\rm host}
\agt 10^{-5}$ ($\overline{m}$ is the orbit-averaged mass of the
subhalos), resolved in the $N$-body simulations.
At relatively small redshifts for both $M_{\rm host} = 10^{13} M_\odot$
and $10^{7} M_\odot$, we find very good agreement between the two
prescriptions.
We also check the applicability of the analytical approach by
comparing the results with those of $N$-body simulations of small-mass
hosts at higher redshift, $z=32$, for which $\eta$ distribution at $z=7$
of Reference.~\cite{Wetzel:2010kz} was adopted.
Even at the very high redshift and for very small host mass of
$M_{\rm host} = 10^{-2}M_\odot$, we still find reasonable agreement
within differences of factor of a few in $\dot{m}$ between results
obtained by the Monte Carlo approaches and the N-body simulations.
Although we cannot test the validity of our Monte Carlo approach for
$\overline m/M_{\rm host} \ll 10^{-5}$ in comparison with the $N$-body
simulations, these agreements that have been seen in
Figure.~\ref{fig:mdotvsm} from very small to large hosts as well as from
very high to low redshifts give us confidence that our analytical
prescription captures physics of tidal stripping, and hence can be
applied even to the cases with an extremely small mass ratio
$\overline{m}/M_{\rm host}$.

From each calculation of $(M_\mathrm{host}, z)$, we fitted the values
of $A$ and $\zeta$ in Eq.~(\ref{eq:mdot}). 
We then derived the dependence of $A$ and $\zeta$ on the host mass
$M_\mathrm{host}$ and $z$ as:
\begin{eqnarray}
\log A&=&\left[-0.0003\log\left(\frac{M_\mathrm{host}}{M_\odot}\right)+0.02\right]z \nonumber\\
&&{}+0.011\log\left(\frac{M_\mathrm{host}}{M_\odot}\right)-0.354, \label{Adependence}\\ 
\zeta&=&\left[0.00012\log\left(\frac{M_\mathrm{host}}{M_\odot}\right)-0.0033\right]z \nonumber\\
&&{}-0.0011\log\left(\frac{M_\mathrm{host}}{M_\odot}\right)+0.026.\label{zetadependence}
\end{eqnarray}
We obtain the relations, Eqs.(\ref{Adependence}) and
(\ref{zetadependence}), from results of the Monte Carlo simulations that
covers the host mass from $M_\mathrm{host}=10^{-6}M_\odot$ to
$10^{16}M_\odot$ and the redshift from $z=0$ to $7$.

\section{Results}
\label{sec:result}

By combining the tidal mass-loss rate (Sec.~\ref{sec:stripping}) with
the analytical prescription for computing density profiles after tidal
stripping as well as the subhalo accretion onto evolving hosts
(Sec.~\ref{sec:subhalostructure}), we are able to calculate quantities
of interest related to the subhalos.
They are the subhalo mass function and the annihilation boost factor,
discussed below in Secs.~\ref{ssec:massfunction} and \ref{ssec:boost},
respectively.

We first fix the reshift of interest $z_0$ and the host mass at that
redshift, $M_0$.
For each set of ($M_0$, $z_0$), we uniformly sample $m_{\rm acc}$ in
logarithmic space between $10^{-6}M_\odot$ and $0.1M_0$, and $z_{\rm
acc}$ between $z_0+0.1$ and $10$.
Each combination is characterized by a subscript $i$, ($\ln m_{{\rm
acc},i}$, $z_{{\rm acc},i}$).
Its weight $w_i$ is chosen to be proportional to the subhalo accretion
rate from the extended Press-Schechter formalism
(Appendix~\ref{ssec:Subhalo accretion rate}):
\begin{equation}
 w_i \propto \left(\frac{d^2N_{\rm sh}}{d\ln m_{\rm acc}dz_{\rm
		     acc}}\right)_i.
\end{equation}
This weight is normalized such that
\begin{equation}
 \sum_i w_i = N_{\rm sh, total},
\end{equation}
where $N_{\rm sh, total}$ represents the total number of subhalos ever
accreted on the given host by the time $z = z_0$.
It is obtained by numerically integrating $d^2N_{\rm sh}/(d\ln m_{\rm
acc}dz_{\rm acc})$ [Eq.~(\ref{maccdistribution})] over $\ln m_{\rm acc}$
and $z_{\rm acc}$.
This way, we essentially approximate the integral of the distribution of
$\ln m_{\rm acc}$ and $z_{\rm acc}$ as
\begin{equation}
 \int d\ln m_{\rm acc} \int dz_{\rm acc} \frac{d^2N_{\rm sh}}{d\ln
  m_{\rm acc}dz_{\rm acc}} ~\to ~\sum_i w_i.
\end{equation}

\subsection{Mass function of subhalos}
\label{ssec:massfunction}

As discussed in Sec.~\ref{AM}, the subhalo mass at $z_0$ after tidal
stripping, $m_{0,i}$, is calculated by integrating Eq.~(\ref{eq:mdot})
over cosmic time from that corresponding to $z = z_{{\rm acc},i}$ to $z
= z_0$.
The parameters $A$ and $\zeta$ are taken from Eqs.~(\ref{Adependence})
and (\ref{zetadependence}), respectively.
For each $i$, we obtain the subhalo concentrations at accretion
following the log-normal distribution $P(c_{\rm vir,acc}|m_{{\rm
acc},i},z_{{\rm acc},i})$ as discussed in
Sec.~\ref{sec:subhalostructure} and calculate the scale radius $r_{s,i}$
and characteristic density $\rho_{s,i}$ at redshift
$z_{\mathrm{acc},i}$, as functions of $c_{\rm vir,acc}$. 
Those quantities after tidal stripping is then obtained from those
before the stripping combined with the stripped mass $m_{0,i}$, as in
Sec.~\ref{sec:subhalostructure}.
If the truncation radius, $r_{t,i}$, is found smaller than $0.77r_{s,i}$
at $z = z_0$ after the tidal stripping, we exclude the subhalo from
calculation of the mass function as it is regarded as completely
disrupted.

The subhalo mass function is then constructed as the distribution of
$m_{0,i}$ properly weighted by $w_i$ with the condition of tidal
disruption as follows: 
\begin{eqnarray}
 \frac{dN_{\rm sh}}{dm} &=& \sum_i w_i \delta(m - m_{0,i})
  \nonumber\\&&{}\times
  \int dc_{\rm vir,acc}P(c_{\rm vir,acc}|m_{{\rm acc},i},z_{{\rm
  acc},i})
  \nonumber\\&&{}\times
  \Theta [r_{t,i}(z_0|c_{\rm vir,acc})-0.77r_{s,i}(z_0|c_{\rm
  vir,acc})],
  \nonumber\\
\label{eq:SHMF}
\end{eqnarray}
where $\delta(x)$ and $\Theta(x)$ are the Dirac delta function and
Heaviside step function, respectively.

\begin{figure}
\begin{center}
\includegraphics[width=7.5cm]{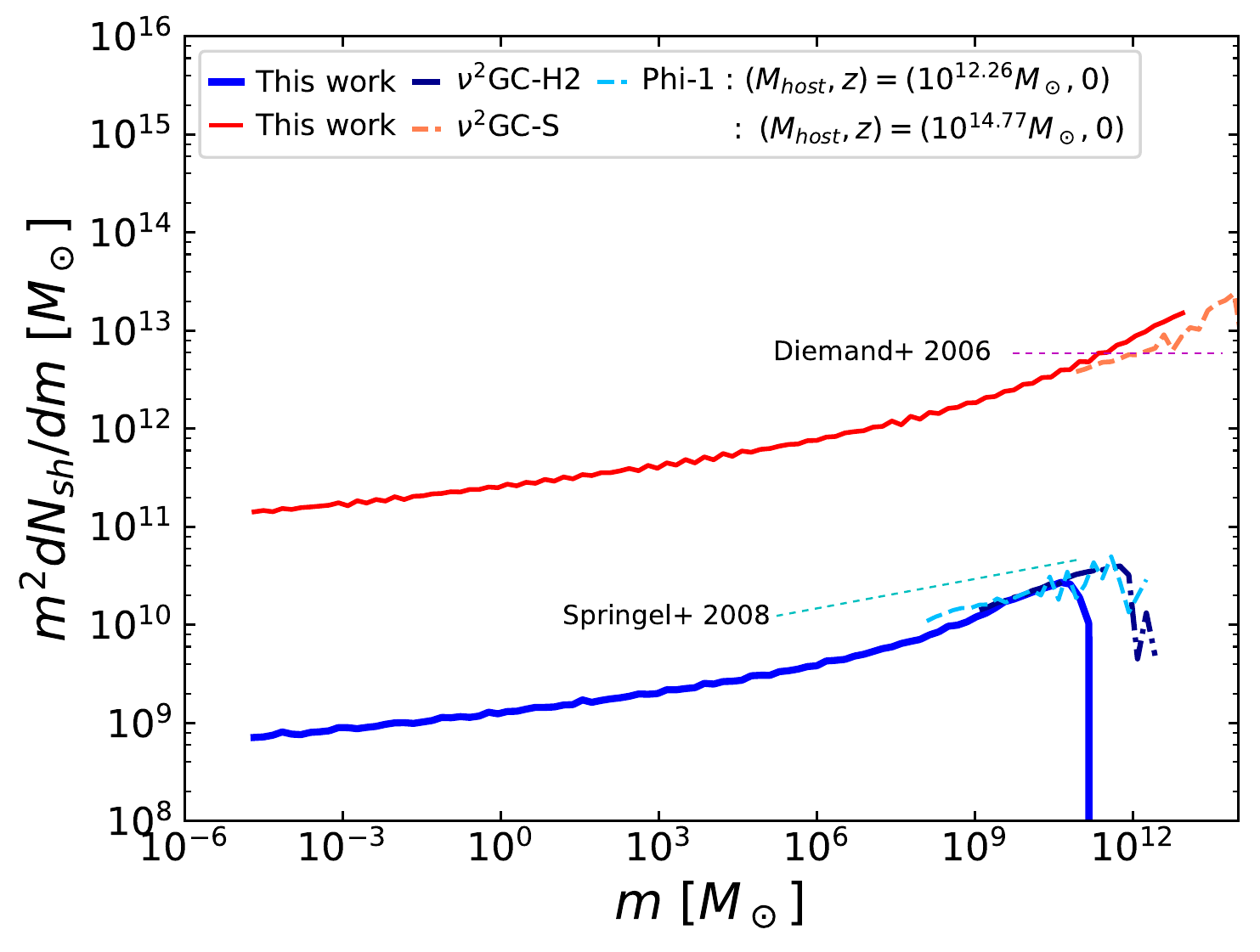}
\includegraphics[width=7.5cm]{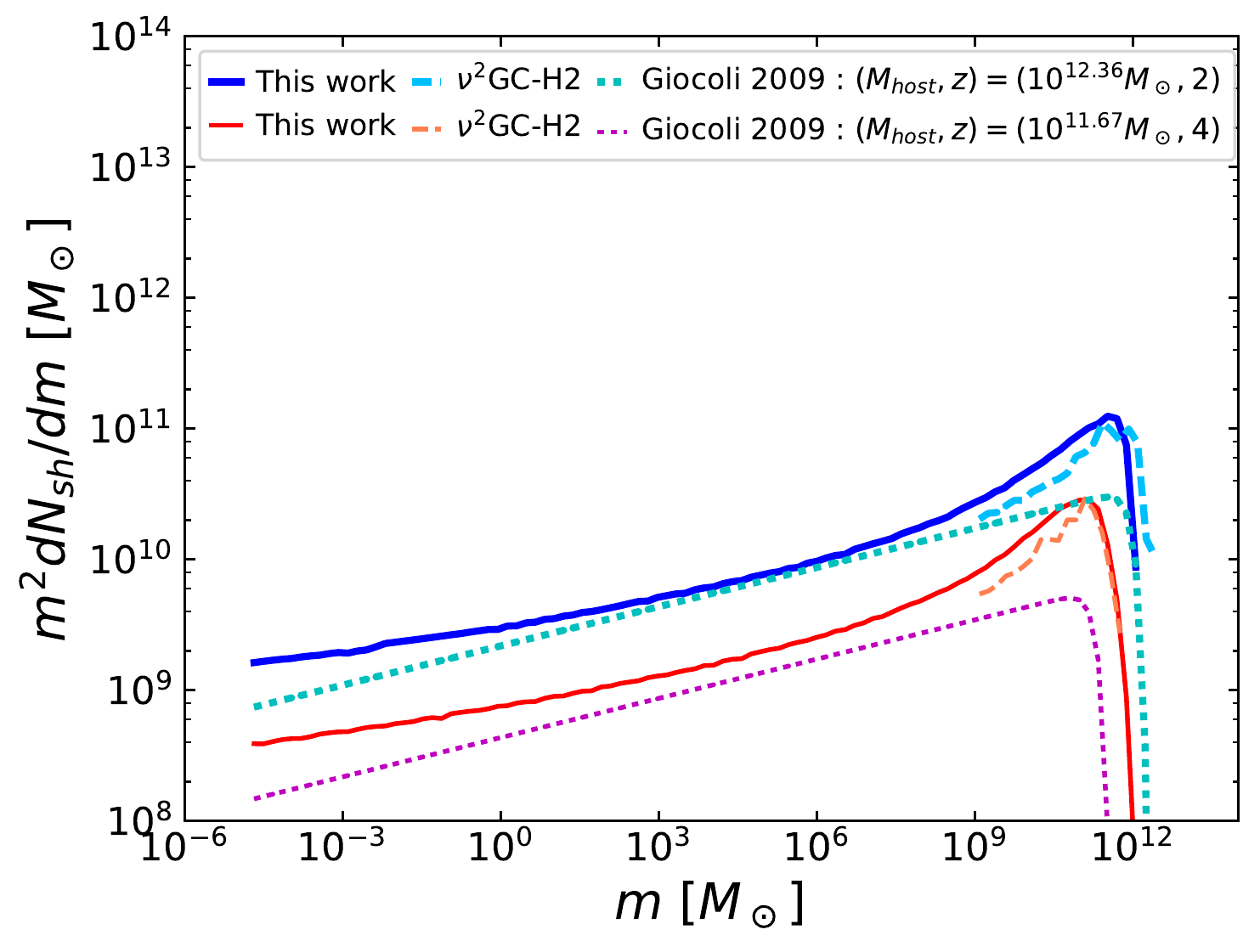}
\includegraphics[width=7.5cm]{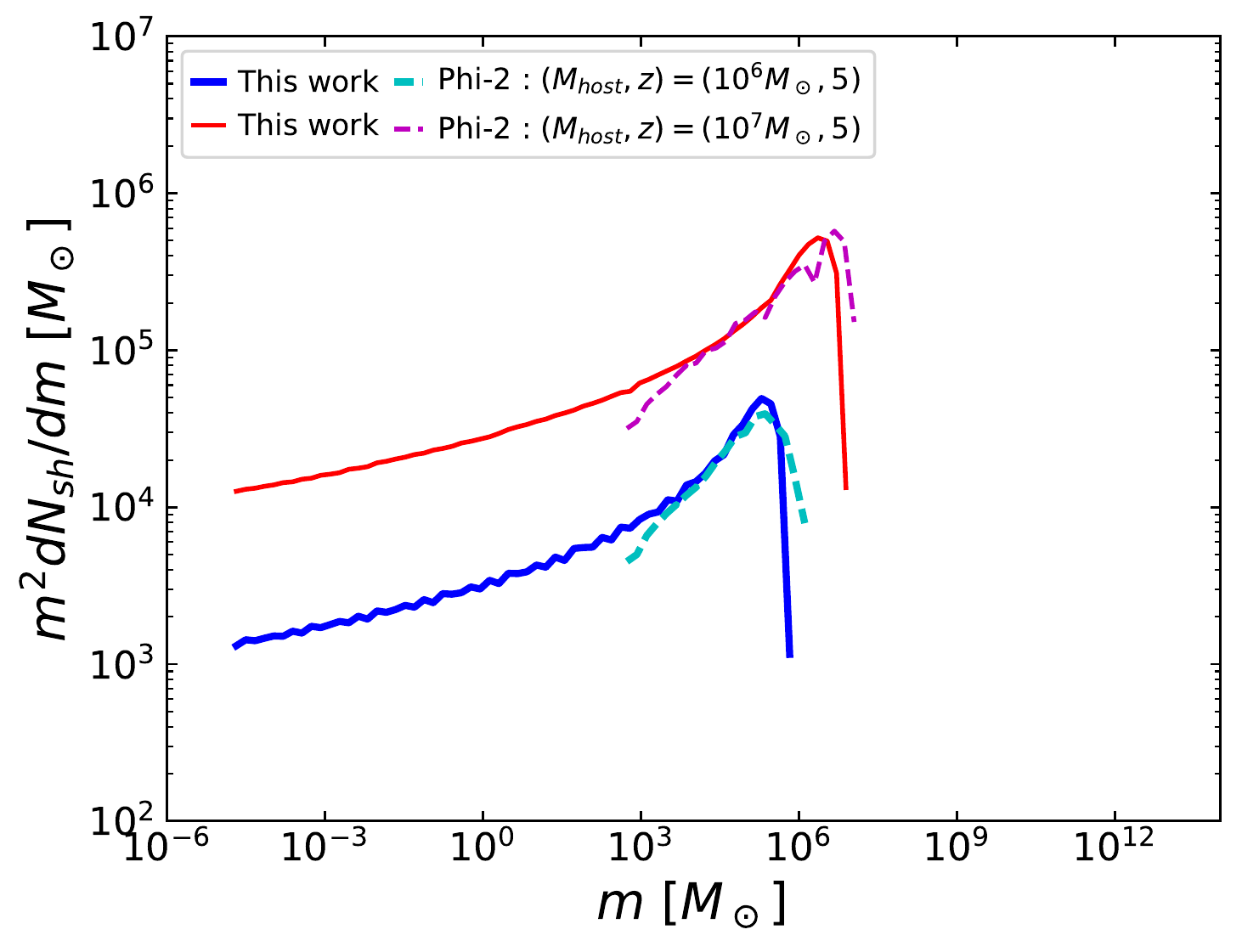}
\caption{Mass function of subhalos and comparison with the results of
 numerical simulations. {\it Top}: Comparison at $z=0$. Thick (blue) lines correspond to the case of $M_\mathrm{host}=1.8\times
 10^{12}M_\odot$ while thin (red) lines to
 $5.9\times10^{14}M_\odot$. Solid lines show the mass function obtained
 in our analytical modelings and dashed lines show those obtained by the
 N-body simulations in Table~\ref{tab:simulations}. Fitting fnctions in
 Reference.~\cite{Springel:2008cc} for
 $M_\mathrm{host}=1.8\times10^{12}M_\odot$ and in
 Reference.~\cite{Diemand:2006ey} for $5.9\times 10^{14}M_{\odot}$ are also
 shown for comparison. {\it
 Middle}: Cases of $M_\mathrm{host}=2.3\times10^{12}M_\odot$ at $z=2$ (solid, blue) and $M_\mathrm{host}=4.7\times10^{11}M_\odot$ at
 $z=4$ (thin, red) in comparison again with the simulations in
 Table~\ref{tab:simulations} and Reference.~\cite{Giocoli:2009ie}. {\it
 Bottom}: Comparison at $z=5$ for the cases of
 $M_\mathrm{host}=10^6M_\odot$ (solid, blue lines) and $10^7M_\odot$
 (thin, red lines) with the Phi-2 simulations. Note that some of the
 lines corresponding to our $N$-body simulations extends toward large
 masses, because halos of various masses around a given geometric mean
 have been stacked in order to derive the mass functions.}
\label{fig:shmf}
\end{center}
\end{figure}

The subhalo mass function has been studied most commonly through $N$-body
simulations in the literature.
We show $m^2 dN_{\rm sh}/dm$ obtained by the numerical simulations and
by our analytical model [Eq.~(\ref{eq:SHMF})] in Figure.~\ref{fig:shmf}.
In the top panel of Figure.~\ref{fig:shmf}, we compare the subahalo mass
function for host masses $M_\mathrm{host}=1.8\times10^{12}M_\odot$ and
$5.9\times10^{14}M_\odot$ at $z=0$ with the fitting functions to the
results of References.~\cite{Springel:2008cc} and \cite{Diemand:2006ey},
respectively.
In both cases, the simulations and analytical models show reasonable
agreement, while our model predicts fewer subhalos.
We also show the results of $\nu^2$GC-S, $\nu^2$GC-H2, and Phi-1
simulations, all of which show better agreement with our analytical
results.
In the middle panel of Figure.~\ref{fig:shmf}, we compare the mass function
at $z=2$ and $z=4$ with results of Reference.~\cite{Giocoli:2009ie} as well as
$\nu^2$GC-H2, for the host that has the mass of
$M_\mathrm{host}=10^{13}M_\odot$ at $z=0$.
This again shows very good agreement between the two approaches, where
the subhalos are resolved in the numerical simulations.
Our model can also be applied to cases of even smaller hosts.
In the bottom panel of Figure.~\ref{fig:shmf}, we compare the subhalo mass
function for $M_\mathrm{host}=10^6M_\odot$ and $10^7M_\odot$ at $z = 5$
with the results of the Phi-2 simulations.
Down to the resolution limit of the simulations that are around
500--$1000M_\odot$, both the calculations agree well.
Hence, the subhalo mass functions from our analytical model is well
calibrated to the results of the numerical simulations at high masses,
and since it is physically motivated, the behavior at low-mass end down
to very small masses can also be regarded as reliable.

\begin{figure}
 \begin{center}
  \includegraphics[width=7.5cm]{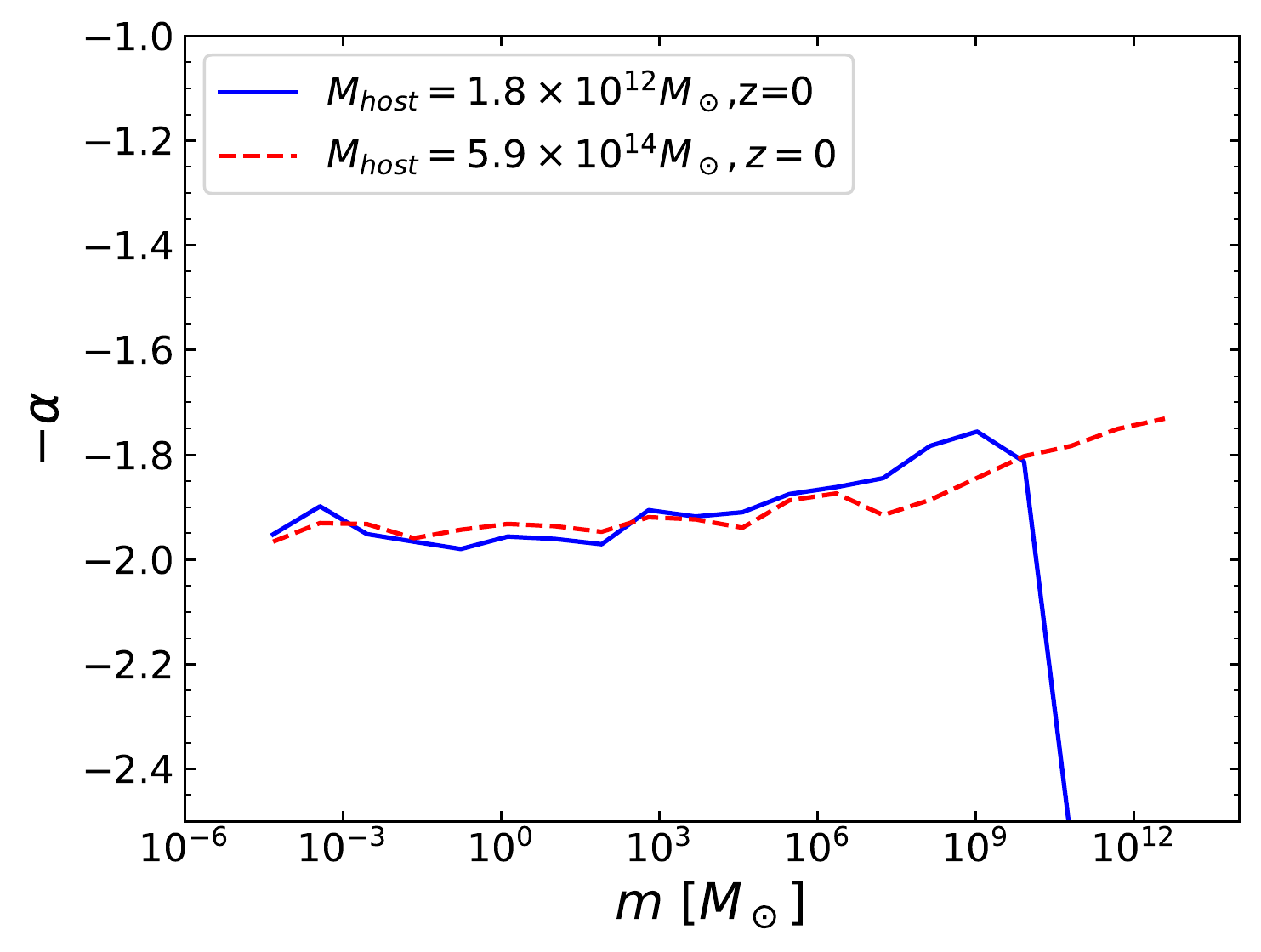} 
  \includegraphics[width=7.5cm]{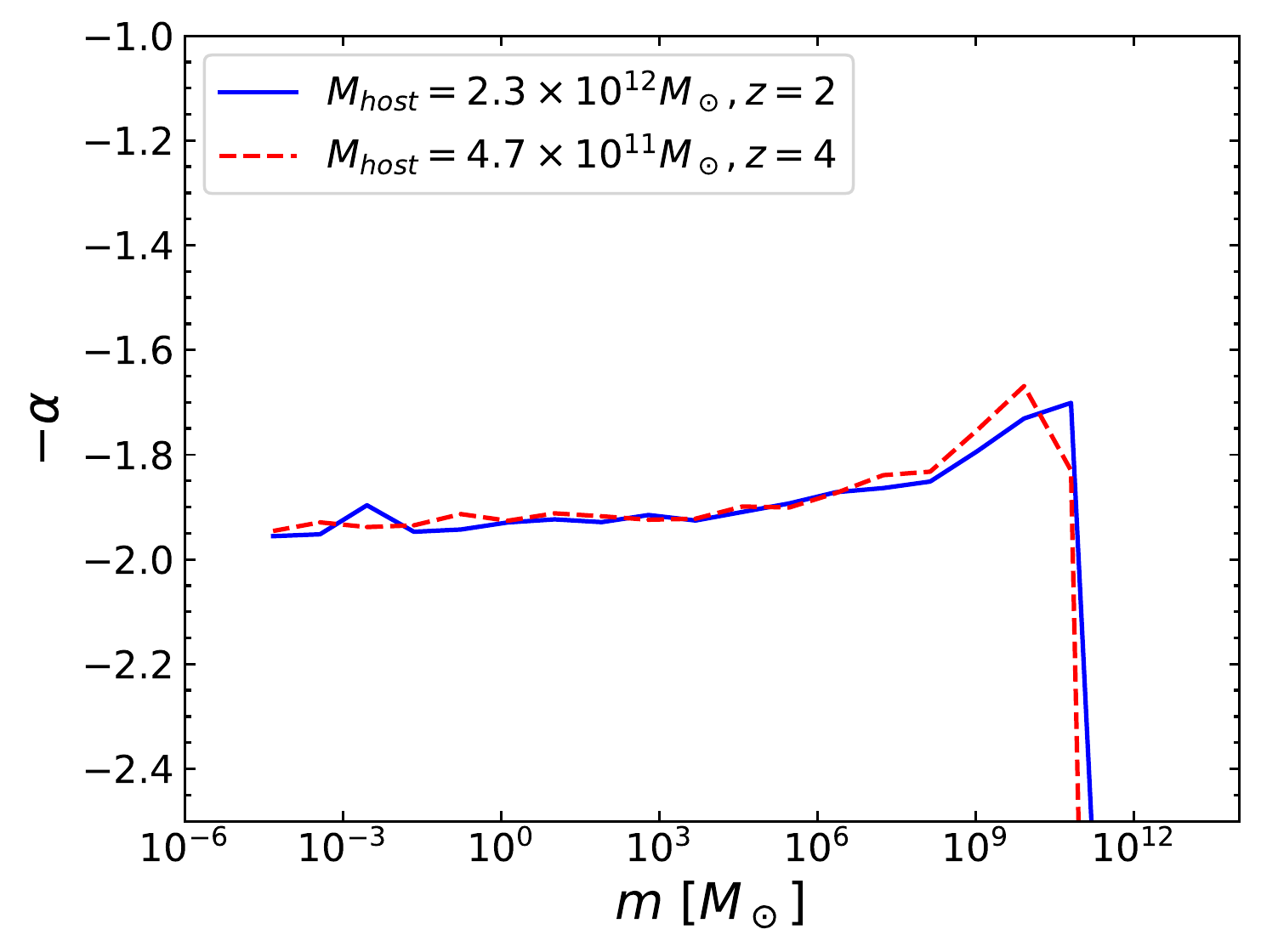}
  \includegraphics[width=7.5cm]{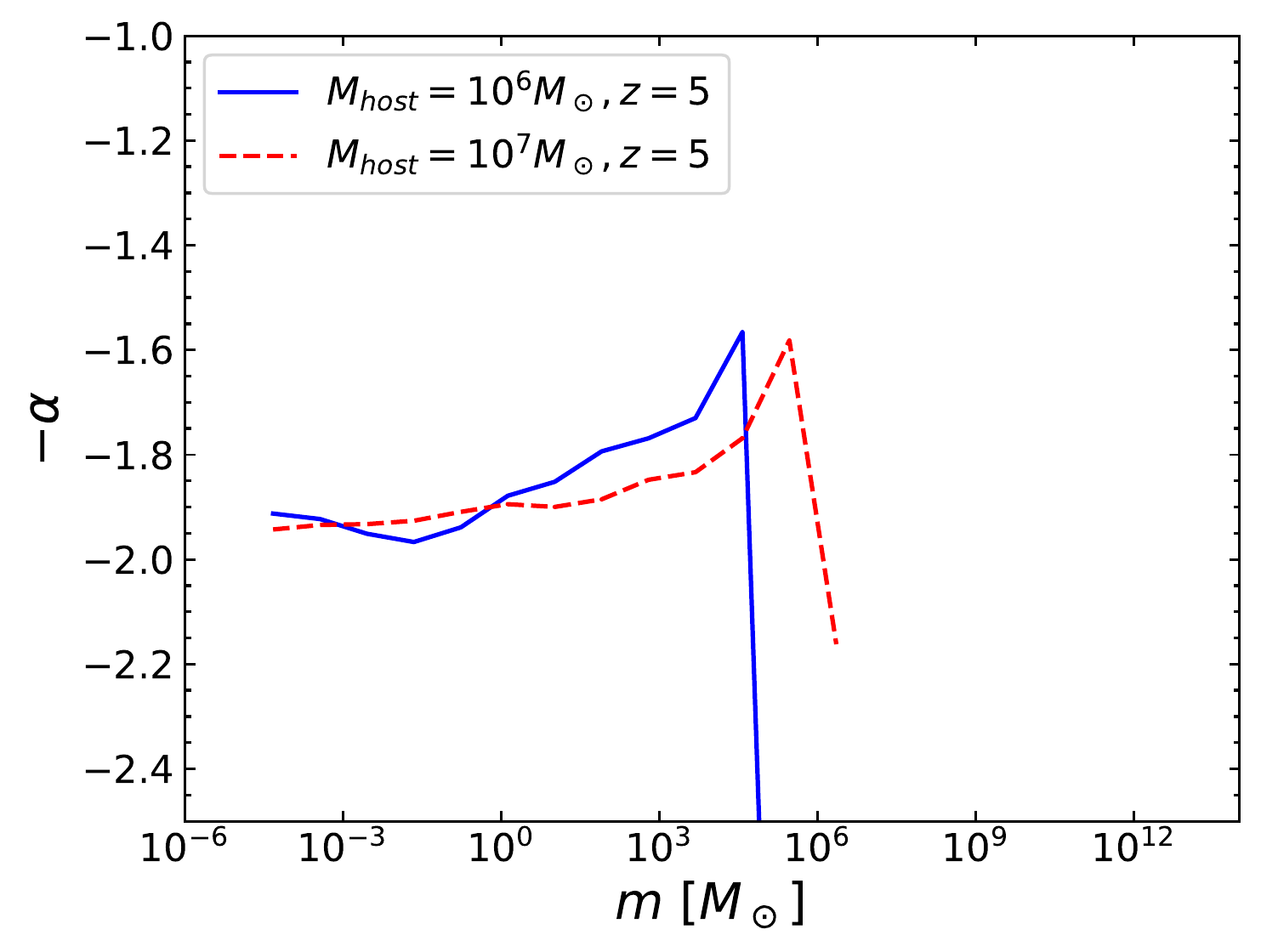}
  \caption{The slope of the subhalo mass function $-\alpha = d\ln
  (dN_{\rm sh}/dm)/d\ln m$ as a function of $m$. The slope was averaged
  over mass bins of width $\Delta \log m = 1$. }
  \label{fig:slope}
 \end{center}
\end{figure}

In Figure.~\ref{fig:slope}, we show the slope of the subhalo mass function
\begin{equation}
 -\alpha = \frac{d\ln (dN_{\rm sh}/dm)}{d\ln m},
\end{equation}
(i.e., $dN_{\rm sh}/ dm \propto m^{-\alpha}$) for the same models as in
Figure.~\ref{fig:shmf}.
We find that the slope lies in a range between $-2$ and $-1.8$ for a
large range of $m$ except for lower and higher edges where the mass
function features cutoffs.
This is consistent with one of the findings from the numerical
simulations, again confirming validity of our analytical model.

\begin{figure}
 \begin{center}
  \includegraphics[width=8.5cm]{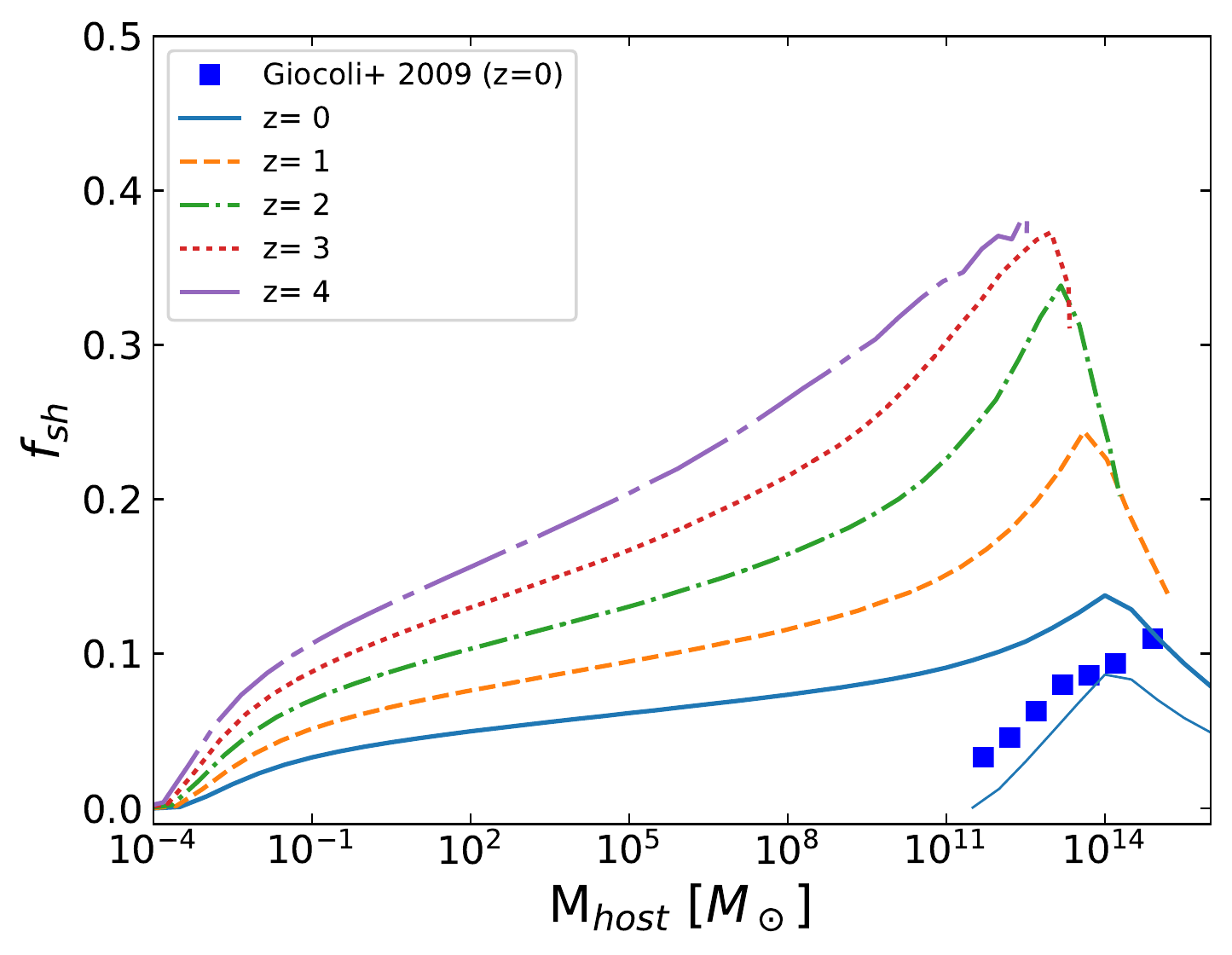}
  \caption{Mass fraction of the host halo in the form of subhalos,
  $f_{\rm sh}$ as a function of $M_{\rm host}$, for $z = 0$, 1, 2, 3,
  and 4. Blue squares represent the subhalo mass fractions in
  Reference.~\cite{Giocoli:2009ie}, which are derived using subhalos with
  masses between $1.73\times10^{10}h^{-1}M_\odot$ and
  $0.1 M_\mathrm{host}$. Solid thin line shows the corresponding subhalo
  mass fraction in our calculation.}
  \label{fig:fsub}
 \end{center}
\end{figure}

Figure~\ref{fig:fsub} shows the mass fraction of the host mass that is
contained in the form of the subhalos:
\begin{equation}
 f_{\rm sh} =  \frac{1}{M_{\rm host}}\int_{10^{-6}M_\odot}^{0.1 M_{\rm
  host}} dm\ m \frac{dN_{\rm sh}}{dm}.
\end{equation}
At $z = 0$, this fraction is smaller than $\sim$10\% level up to
cluster-size halos.
We also find that $f_{\rm sh}$ is larger for higher redshifts, as the
effect of tidal mass loss is suppressed compared with the case of $z =
0$.
In Figure.~\ref{fig:fsub}, we also show the results of $N$-body simulations by
Reference.~\cite{Giocoli:2009ie} for the subhalo mass fraction between
$1.73\times10^{10}h^{-1}M_\odot$ and $0.1 M_\mathrm{host}$, which is in
good agreement with our analytical result for the same quantity.

\subsection{Subhalo boost}
\label{ssec:boost}

\subsubsection{Case of smooth subhalos}

The gamma-ray luminosity from dark matter annihilation in the smooth NFW
component of the host halo with mass $M$ and redshift $z$ is obtained as
\begin{equation}
 L_{\rm host}(M) \propto \int dc_{\rm vir} P(c_{\rm
  vir}|M,z)\rho_s^2r_s^3 \left[1-\frac{1}{(1+c_{\rm vir})^3}\right],
  \label{eq:Lhost}
\end{equation}
where $P(c_{\rm vir}|M,z)$ is again the log-normal distribution of the
host's concentration parameter given $M$ and $z$, and the scale radius
$r_s$ and the characteristic density $\rho_s$ are both dependent on
$c_{\rm vir}$ as well as on $M$ and $z$.
The constant of proportionality of this relation includes particle
physics parameters such as the mass and annihilation cross section of
dark matter particles, but since here we are interested in the ratio of
the luminosities between the subhalos and the host, their dependence
cancels out.

Subhalo boost factor quantifies the contribution of all the subhalos to
the total annihilation yields compared with the contribution from the
host.
It is defined as
\begin{equation}
 B_{\rm sh}(M) = \frac{L_{\rm sh}^{\rm total}(M)}{L_{\rm host}(M)},
  \label{eq:boost}
\end{equation}
such that the total luminosity from the halo is given as $L_{\rm total}
 = (1+B_{\rm  sh})L_{\rm host}$.
The luminosity from a single subhalo $i$ characterized with its
accretion mass $m_{{\rm acc},i}$ and redshift $z_{{\rm acc},i}$, as well
as its virial concentration $c_{\rm vir,acc}$ is
\begin{equation}
 L_{{\rm sh},i} \propto \rho_{s,i}^2r_{s,i}^3
  \left[1-\frac{1}{(1+r_{t,i}/r_{s,i})^3}\right],
  \label{eq:Lsh}
\end{equation}
where $r_{s,i}$, $r_{t,i}$, and $\rho_{s,i}$ are the scale radius,
truncation radius, and characteristic density of the subhalo $i$ after
it experienced the tidal mass loss, and hence they are functions of
$m_{{\rm acc},i}$, $z_{{\rm acc},i}$, and $c_{\rm vir,acc}$ as well as
the mass of the host $M$ and redshift $z$
(Sec.~\ref{sec:subhalostructure}).
The total subhalo luminosity $L_{\rm sh}^{\rm total}(M)$ is then
obtained as the sum of $L_{{\rm sh},i}$ with weight $w_i$ and averaged
over $c_{\rm vir,acc}$ with its distribution:
\begin{eqnarray}
 L_{\rm sh}^{\rm total}(M) &=& \sum_i w_i
  \int dc_{\rm vir,acc}P(c_{\rm vir,acc}|m_{{\rm acc},i},z_{{\rm acc},i})
  \nonumber\\&&{}\times
  L_{{\rm sh},i}(z|c_{\rm vir,acc})
  \nonumber\\&&{}\times
  \Theta [r_{t,i}(z|c_{\rm vir,acc})-0.77r_{s,i}(z|c_{\rm
  vir,acc})].
  \nonumber\\  \label{eq:Lsh_total}
\end{eqnarray}

\subsubsection{Presence of sub-subhalos}

The discussions above, especially Eq.~(\ref{eq:Lsh}), are based on the
assumption that the density profile of subhalos is given by smooth NFW
function.
Subhalos, however, contain their own subhalos: i.e., sub-subhalos,
which again contain sub-sub-subhalos, and so on.
This is because the subhalos, before accreting onto their host, were
formed by mergers and accretion of even smaller halos.
In the following, we refer to them as sub$^n$-subhalos; the discussion
above correspond to the case of $n = 0$, where subhalos do not include
sub-subhalos.

We include the effect of sub$^n$-subhalos iteratively.
In the case of $n \ge 1$, when a subhalo $i$ accretes at $z_{{\rm acc},
i}$ with a mass $m_{{\rm acc}, i}$, we give it a sub-subhalo boost
$B_{\rm sh}^{(n-1)}(m_{{\rm acc},i}, z_{{\rm acc},i})$ obtained from the
previous iteration; for $n = 1$, it is Eq.~(\ref{eq:boost}) evaluated at
$m_{{\rm acc},i}$ and $z_{{\rm acc},i}$.
After the subhalo exprience the mass loss, its sub-subhalos as well as
smooth component are stripped away up to the tidal radius $r_{t,i}$.
Since the sub-subhalo distribution (that the gamma-ray brightness
profile from the sub-subhalos follows) is flatter than the brightness
profile of the subhalo's smooth component that is proportional to the
NFW profile squared, the sub-subhalo boost decreases.
In order to quantify this effect, we assume that the sub-subhalos are
distributed as $n_{\rm ssh}(r) \propto (r^2+r_s^2)^{-3/2}$ (see, e.g.,
Reference.~\cite{Ando:2013ff} and references therein), and further
assuming that $r_s$ and $\rho_s$ hardly change after mass loss, the
total sub-subhalo luminosity enclosed within $r$ is 
\begin{equation}
 L_{{\rm ssh},i}(<r)
\propto \ln
\left[\sqrt{1+\left(\frac{r}{r_{s,i}}\right)^2}+\frac{r}{r_{s,i}}\right]-\frac{r}{\sqrt{r^2+r_{s,i}^2}}.
\end{equation}
On the other hand, the enclosed luminosity from the smooth NFW component
is 
\begin{equation}
 L_{{\rm sh},i}(<r) \propto 1-\left(1+\frac{r}{r_{s,i}}\right)^{-3}.
\end{equation}
The sub-subhalo boost for the subhalo $i$ at redshift $z$ after $n$-th
iteration is therefore estimated as
\begin{eqnarray}
 B_{{\rm ssh},i}^{(n)}(z) &=& B_{\rm sh}^{(n-1)}(m_{{\rm
  acc},i},z_{{\rm acc},i})
  \nonumber\\&&{}\times
  \frac{L_{{\rm ssh},i}(<r_{t,i})/L_{{\rm ssh},i}(<r_{{\rm
  vir},i})}{L_{{\rm sh},i}(<r_{t,i})/L_{{\rm sh},i}(<r_{{\rm vir},i})},
 \label{eq:Bssh}
\end{eqnarray}
where $r_{{\rm vir},i}$ is the virial radius of the subhalo $i$ at
accretion.

\begin{figure}
\begin{center}
\includegraphics[width=8.5cm]{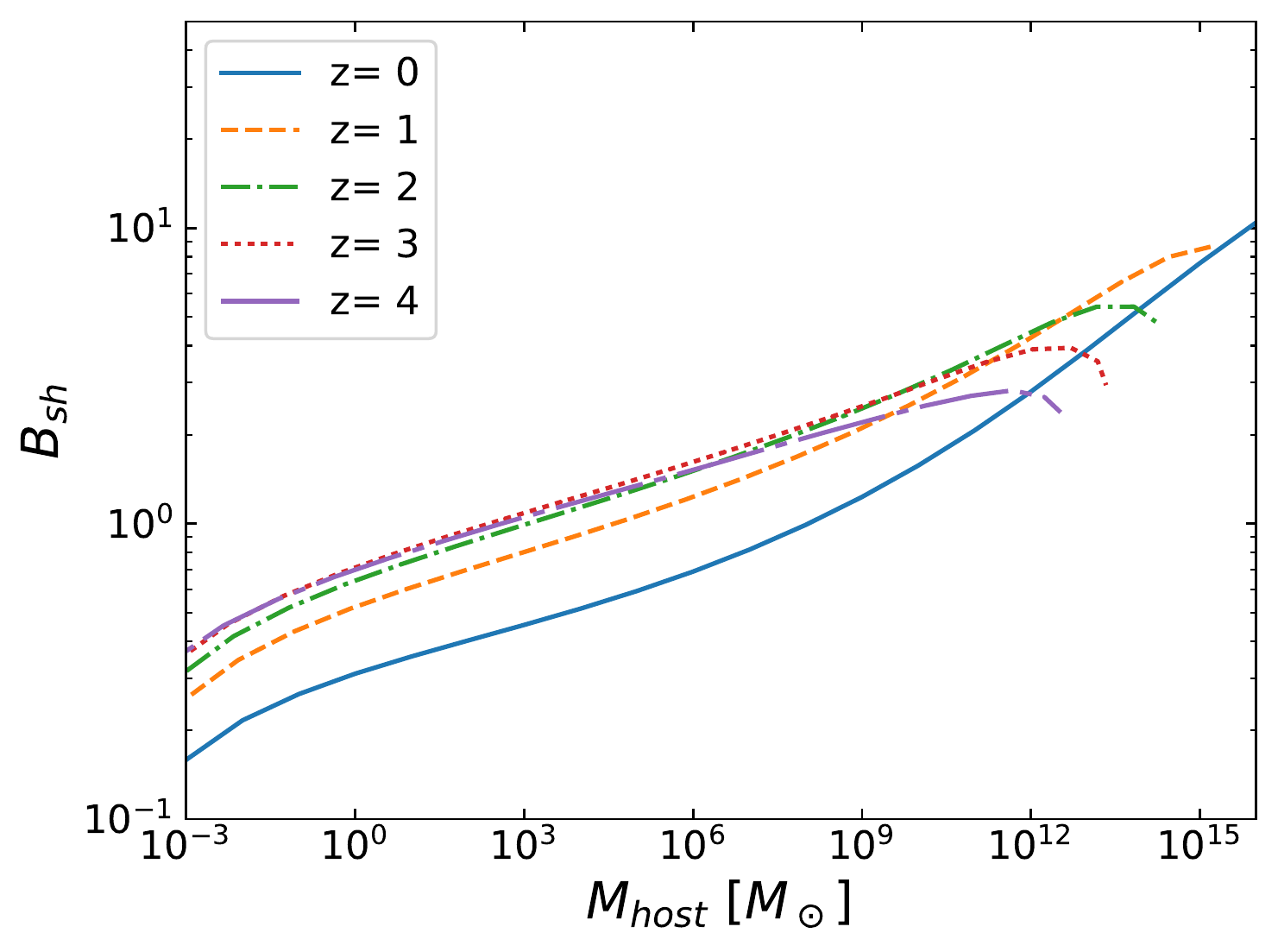}
\caption{Boost factor $B_{\rm sh}=L_{\rm sh}^{\rm total}/L_{\rm host}$
 as a function of the host mass
 $M_\mathrm{host}$ (defined as $M_{200}$) between $10^{-3}M_\odot$ and
 $10^{16}M_\odot$ at observation redshifts $z= 0$, 1, 2, 3, and 4. The
 calculations include up to sub$^3$-subhalos.}
\label{fig:boostz}
\end{center}
\end{figure}

We finally obtain the subhalo boost factor after $n$-th interation
(that takes up to sub$^{n-1}$-subhalos into account), $B_{\rm
sh}^{(n)}(M,z)$, by combining
Eqs.~(\ref{eq:Lhost})--(\ref{eq:Lsh_total}), but also by
multiplying $L_{{\rm sh},i}$ in Eq.~(\ref{eq:Lsh}) with $1+B_{{\rm
ssh},i}^{(n)}(z_0)$ [Eq.~(\ref{eq:Bssh})].
In this calculation, we consider the subhalos accreted after $z=10$,
which assures that we can follow the mass-loss of the subhalos
contributing to the boost factor at $z<5$.
Figure.~\ref{fig:boostz} shows the boost factor $B_{\rm sh}$ as a
function of host mass $M_{\rm host}$ (defined as $M_{200}$) for several
redshifts, after fourth iteration that takes up to sub$^3$-subhalos into
account.
For $z = 0$, the subhalo boost increases gradually with the mass of the
hosts, and reaches to about a factor of ten for cluster-size halos.
The boost for high redshifts is still significant, being on the order of
one, for wide range of host masses.

\begin{figure}
 \begin{center}
  \includegraphics[width=8.5cm]{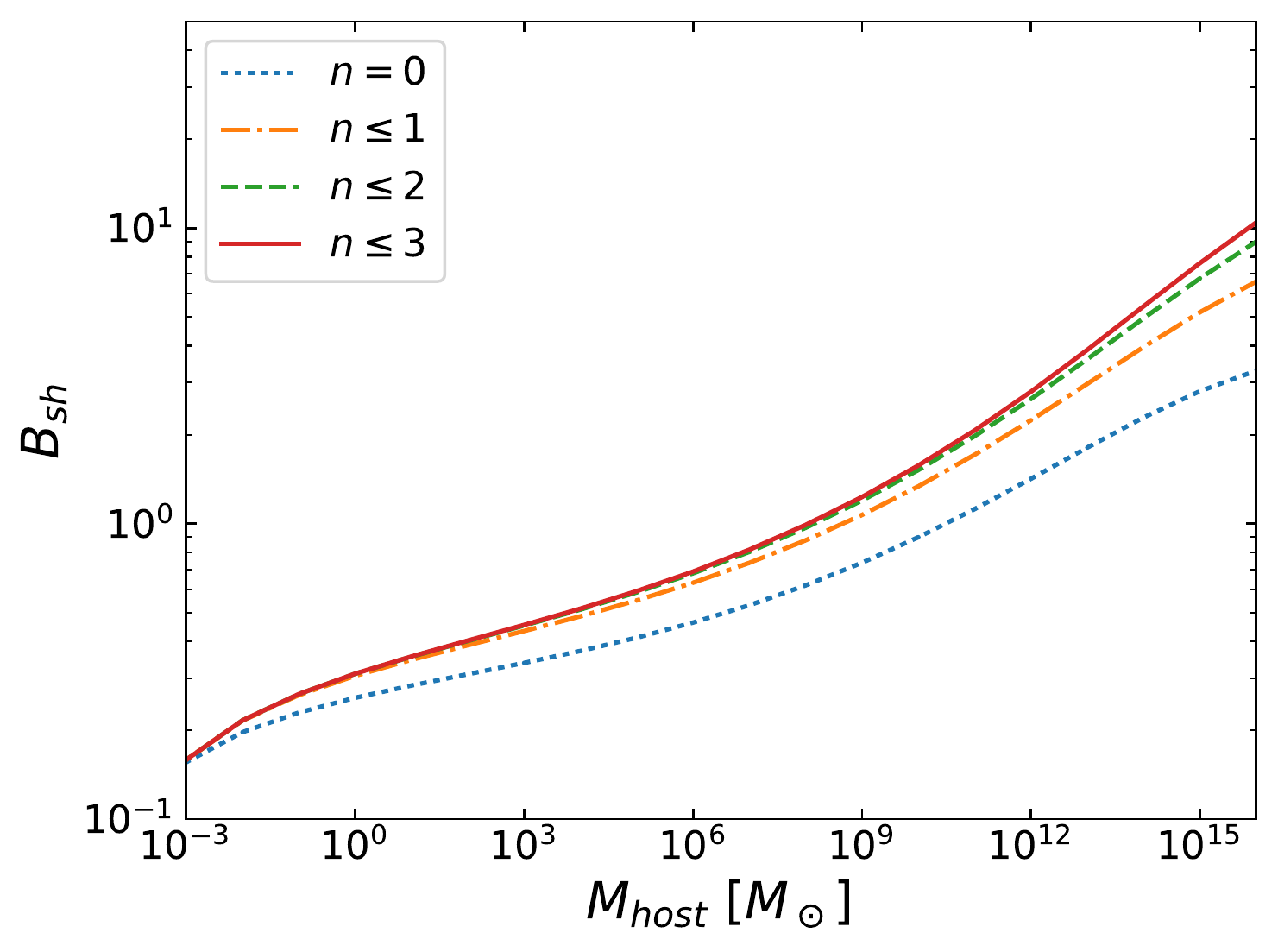}
  \caption{Subhalo boost factor at $z = 0$ including sub$^n$-subhalos;
  i.e., $n$-th sub-substructure.}
  \label{fig:iteration}
 \end{center}
\end{figure}

In Figure.~\ref{fig:iteration}, we investigate the effect of higher-order
substructure: sub$^n$-subhalos.
Including no sub-substructure ($n = 0$) would underestimate the boost by
about a factor of a few for massive host halos such as galaxies and
clusters.
We find that the boost saturates after the third iteration, after which
further enhancement is of several percent level.

\section{Discussion}
\label{sec:discussion}

\subsection{Comparison with earlier work}

The current work updated an analytical model of
Reference.~\cite{Bartels:2015uba}, by (i) implementing the scatter
distribution in the concentration-mass relation for both the host and
subhalos, (ii) calibrating the subhalo mass-loss rate down to extremely
small mass ratio $m/M$ using the Monte Carlo simulations of the tidal
stripping, (iii) extending the calculations of the boost factor as
well as the subhalo mass function beyond $z = 0$, and (iv) including
sub-subhalos and beyond.
They are all essential ingredients to improve the accuracy of the
subhalo modeling, and hence the current work is regarded as direct
update of Reference.~\cite{Bartels:2015uba}.
As the quantitative outcome, we find that the subhalo boost without
contribution from sub-subhalos ($n = 0$) is consistent with the result
of Reference.~\cite{Bartels:2015uba}.
Our result including up to sub$^3$-subhalos further enhances the boost
by a factor of 2--3 for large halos, and extends the calculation down to
$10^{-4} M_\odot$.

The effect of tidal stripping on the annihilation boost has also been
studied in References.~\cite{Zavala:2015ura, Moline:2016pbm} by using
different approaches, but they both have reached a similar conclusion to
that of Reference.~\cite{Bartels:2015uba}.
In particular, Reference.~\cite{Moline:2016pbm} relied directly on $N$-body
simulations to claim that subhalos are more concentrated than field
halos of the equal mass, and hence, the annihilation boost is larger
than previous estimates by, e.g., Reference.~\cite{Sanchez-Conde:2013yxa}.
One of the great advantages of directly using the results from $N$-body
simulations is its accuracy when the discussion concerns the {\it
resolved} regime.
However, each simulation is computationally demanding, and thus,
it is not easy to generalize the discussion to wider ranges of host
masses and redshifts.
In fact, in order to compute the subhalo boost factor as a function of
the host mass, Reference.~\cite{Moline:2016pbm} had to combine the subhalo
concentration-mass relation with the subhalo mass function, for the
latter of which a few phonomenological fitting functions calibrated with
other simulations were adopted.
Hence, the boost factor as its outcome shows a very large range of
uncertainties depending on what model of the mass function one adopts.
In our analytical approach, on the other hand, we are able to perform
physics-based computations of the subhalo boost factor and mass function
in a self-consistent manner, for very wide ranges of masses and
redshifts.

References~\cite{Kamionkowski:2008vw,Kamionkowski:2010mi} developed an
analytical model assuming self-similarity of the substructures, computed
the probability distribution function of the dark matter density that
has a power-law tail, and calibrated it with numerical simulations of
the Galactic halo.
The annihilation boost factor within the volume of the virial radius of
$\sim$200~kpc was found to be $\sim$10, which is slightly larger than
our result.
This, however, agrees with our result based on a different model of
the concentration-mass relation (see Sec.~\ref{ssec:concentration}).

Reference~\cite{Stref:2016uzb} modeled dark matter subhalos in a
Milky-Way-like halo at $z = 0$ by including the effect of the disk
shocking as well as the tidal stripping.
Our result of the annihilation boost factor is consistent with that of
Reference.~\cite{Stref:2016uzb} after integrating over the entire volume of
the halo and assuming the subhalo mass function of $\propto m^{-1.9}$.
Our discussion in Sec.~\ref{sec:stripping} can be expanded to
accommodate the spatial distribution of subhalos, but doing so and
comparing the result with that of Reference.~\cite{Stref:2016uzb} would
include proper modeling of the baryonic component, which is beyond the
scope of the present work.

\subsection{A case without tidal disruption}

Reference~\cite{vandenBosch:2017ynq} recently pointed out that the tidal
disruption for the subhalos with $r_t < 0.77 r_s$ might be a numerical
artifact, and many more subhalos even with much smaller truncation
radius $r_t$ could survive against the tidal disruption.
In this paper, we do not argue for or against the claim of
Reference.~\cite{vandenBosch:2017ynq}, but simply study the implication of
the claim as an optimistic example.
To this end, we repeated the boost calculations without implementing
the constraint $r_t > 0.77 r_s$; i.e., all the subhalos survive no
matter how much mass they lose due to the tidal stripping.
We find that the obtained boost factor hardly changes at any redshift.

\subsection{Dependence on the concentration-mass relation}
\label{ssec:concentration}

\begin{figure}
 \begin{center}
  \includegraphics[width=8.5cm]{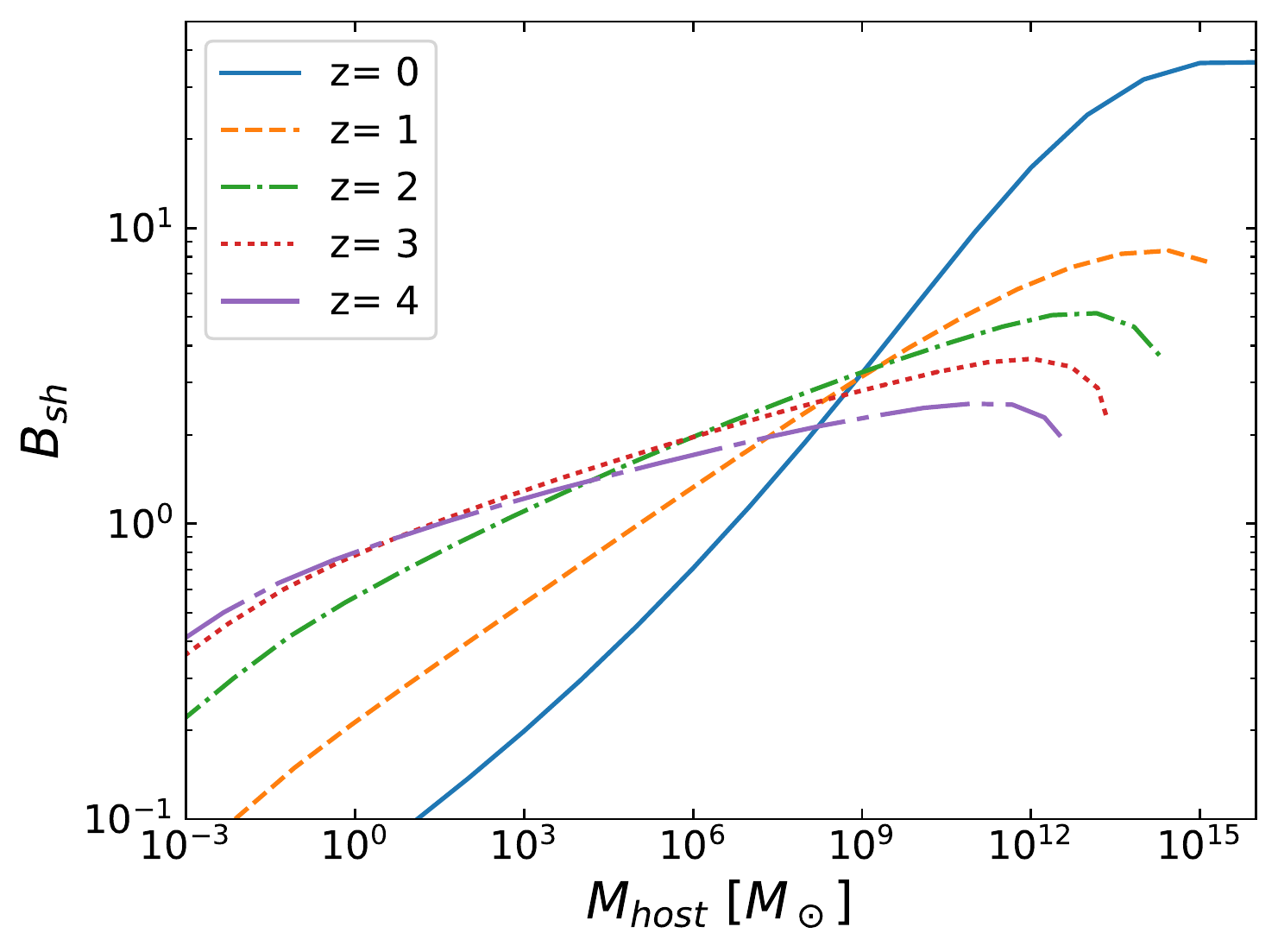}
  \caption{The same as Figure.~\ref{fig:boostz}, but for the concentration-mass relation in Reference.~\cite{Okoli:2015dta}.}
  \label{fig:boostOkoli}
 \end{center}
\end{figure}

In our calculations of the boost factor, we adopted the
mass-concentration relation in Reference.~\cite{Correa:2015dva} as the
canonical model. 
Their derivation is based on the analysis with $N$-body simulations. 
Reference~\cite{Okoli:2015dta} proposed a different concentration-mass
relation based on analytical considerations, which expect higher
concentration especially around $z=0$.
In order to compare the dependence of the boost factor on the different
concentration-mass relations, we also calculated the boost factor
adopting the relation in Reference.~\cite{Okoli:2015dta}. 
In Figure.~\ref{fig:boostOkoli}, we show that the boost factor enhances by
more than a fector of a few if we adopt the concentration-mass relation
of Reference.~\cite{Okoli:2015dta} instead of that of
Reference.~\cite{Correa:2015dva}.
Obtained boost factor directly reflects the difference of the
concentrations at around $z=0$. 
We do not discuss the feasibility of these concentrations since that is
beyond the scope of this paper. 
Our results show that deeper understanding of the concentration-mass
relation is necessary to obtain the boost factor corresponding to the
actual situations.

In Reference.~\cite{Gosenca:2017ybi}, there are some discussions about the
mass-concentration relation and the primordial curvature perturbations
in the early Universe.
If primordial power spectrum has a feature that gives rise to
ultra-compact minihaloes, it may boost dark matter annihilation even
more significantly by changing density profiles and concentration-mass
relation.
Although evaluating the subhalo boost for these specific models is
beyond the scope of our work, we note that such a significant boost
predicted by References.~\cite{Gosenca:2017ybi, Delos:2017thv} may already be
constrained very strongly using the existing gamma-ray data.

\subsection{Contriubtion to the isotropic gamma-ray background}

One of the advantages of our analytical model of the subhalo boost is
capability of calculating the isotropic gamma-ray background (IGRB) from
dark matter annihilation, since we can compute boost factors for
various host masses and the wide range of redshifts, self-consistently.
The intensity of IGRB was most recently measured with
Fermi-LAT~\cite{Ackermann:2014usa}, which was then used to
constrain dark matter annihilation cross
section (e.g.,~\cite{Ackermann:2015tah}).

\begin{figure}
 \begin{center}
  \includegraphics[width=8.5cm]{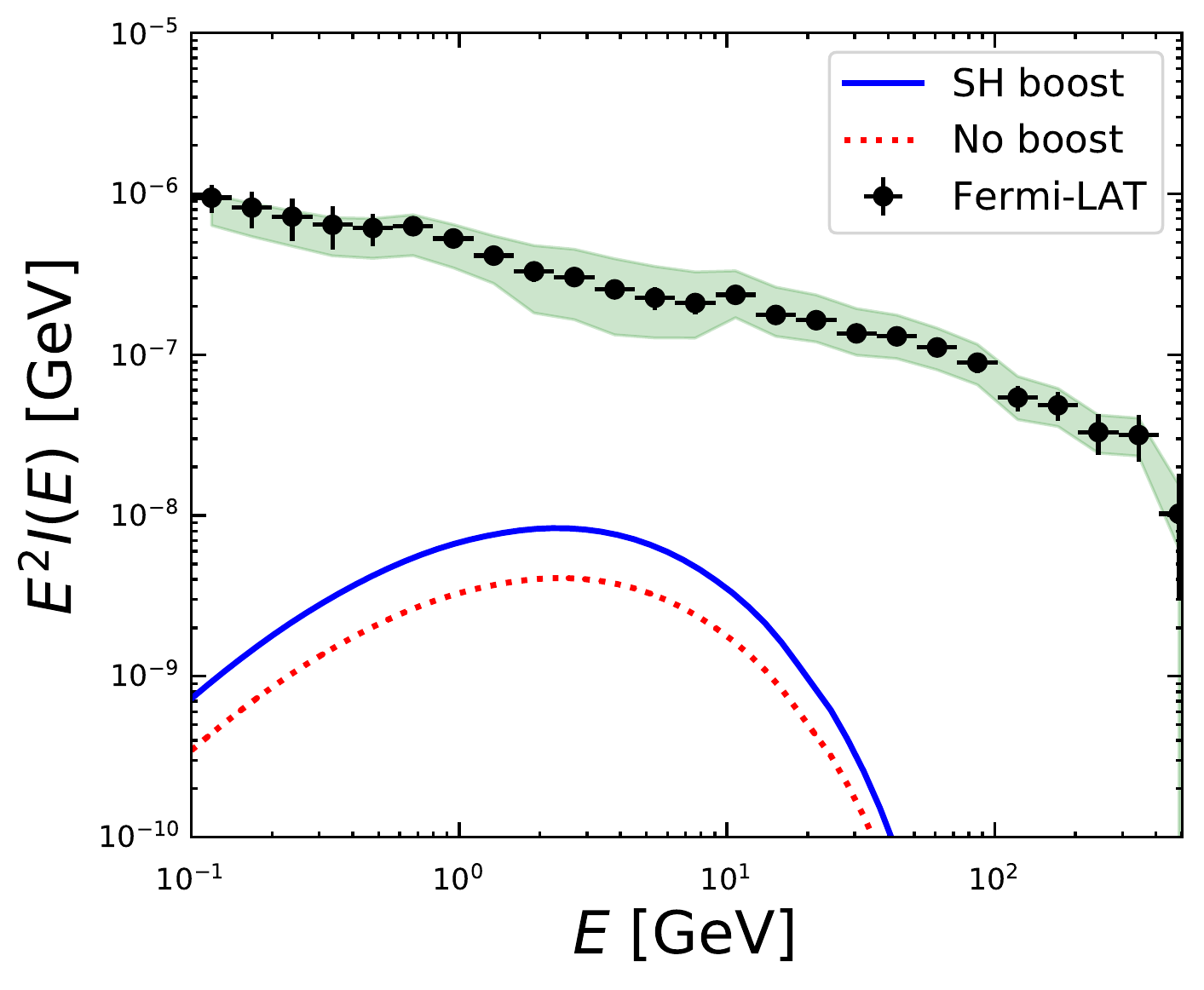}
  \caption{Contriubtion to the IGRB intensity measured by Fermi-LAT from
  dark matter annihilation for $\langle \sigma v \rangle = 2.2\times
  10^{-26}$~cm$^3$~s$^{-1}$, $m_\chi = 100$~GeV, and $b\bar b$ final
  state. The solid (dotted) curve shows the case of the subhalo boost
  (no boost).}
  \label{fig:IGRB}
 \end{center}
\end{figure}

We followed the ``halo model'' approach of Reference.~\cite{Ando:2013ff} to
compute the IGRB contribution from dark matter annihilation, but by
applying the results of the annihialtion boost factor from our
analytical model (Figure.~\ref{fig:boostz}) as well as by including
scatter of the concentration-mass relation.
Figure~\ref{fig:IGRB} shows the IGRB intensity from dark matter
annihilation in the case of the canonical annihilation cross section for
thermal freezeout scenario, $\langle \sigma v \rangle \simeq 2 \times
10^{-26}$~cm$^3$~s$^{-1}$~\cite{Steigman:2012nb}, dark matter mass of
$m_{\chi} = 100$~GeV, and $b\bar b$ final state of the annihilation
($\chi\chi \to b\bar b$).
Our boost model enhances the IGRB intensity by a factor of a few
compared with the case of no subhalo boost.
Note that contribution from the Galactic subhalos (e.g.,
\cite{Ando:2009fp}) is not included, and hence our estimate is
conservative.

\begin{figure}
 \begin{center}
  \includegraphics[width=8.5cm]{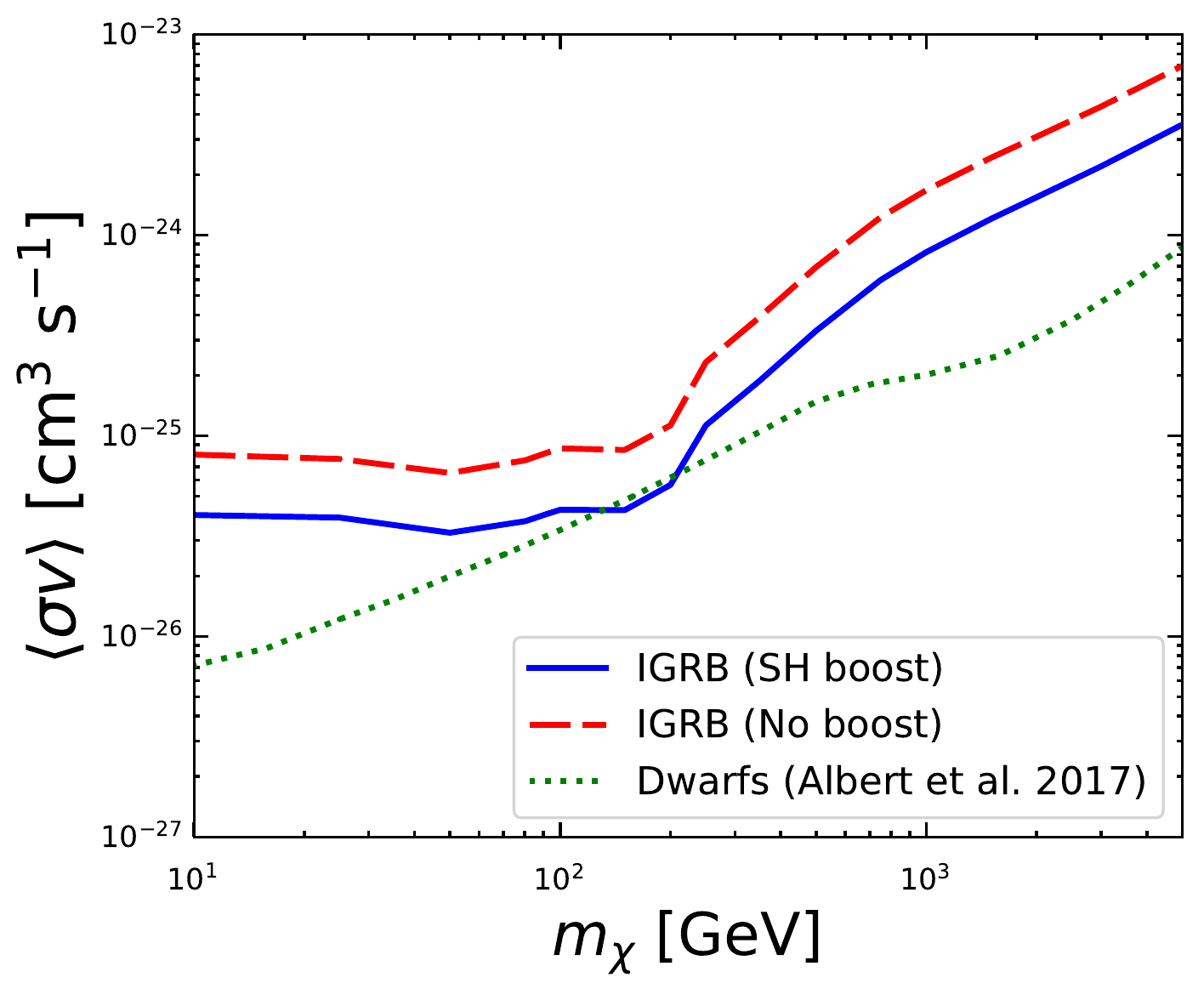}
  \caption{Upper limits on dark matter annihilation cross section at
  95\% confidence level as a function of dark matter mass for $b\bar b$
  final state. Solid and dashed curves are for the canonical boost model
  and without subhalo boost, respectively. For comparison, the result of
  the latest joint-likelihood analysis of 41
  dwarfs~\cite{Fermi-LAT:2016uux} are shown as a dotted curve.}
  \label{fig:sigmav_bb}
 \end{center}
\end{figure}

We then performed a simple analysis of the Fermi-LAT IGRB
data~\cite{Ackermann:2015tah}.
We included two components: (1) dark matter annihilation of a given mass
$m_\chi$ and assuming a $b\bar b$ final states, and (2) an
``astrophysical'' power-law component with a cutoff, for which we adopt
the best-fit spectral shape, $I_{\rm astro}(E) \propto
E^{-2.32}\exp(-E/279~{\rm GeV})$~\cite{Ackermann:2015tah}.
By adopting normalizations of these components as two free parameters
for the fit, we performed a $\chi^2$ analysis in order to obtain the
upper limits on $\langle \sigma v\rangle$.
For the IGRB data, we adopt those for a foreground model ``A'' in
Reference.~\cite{Ackermann:2015tah}, but treat statistical and systematic
uncertainties as independent errors.
Figure~\ref{fig:sigmav_bb} shows the upper limits on $\langle \sigma
v\rangle$ at 95\% confidence level ($\Delta\chi^2 = 2.71$) using our
canonical boost model as well as the case of no boost.
Our updated boost model improves the limits by a factor of a few nearly
indepently of dark matter mass (see also, e.g.,
References.~\cite{Cholis:2013lwa, DiMauro:2015tfa} for earlier results).
This enhancement is calculated consistently as our formalism
automatically computes all the subhalo properties at once including mass
function and the boost factor.
We also compare our limits with the latest results of the joint
likelihood analysis of 41 dwarf spheroidal
galaxies~\cite{Fermi-LAT:2016uux}, which set the benchmark as the most
robust constraints on dark matter annihilation.

Although some improvements of the limit obtained from the observations
of dwarf spheroidal galaxies also can be expected, we conservatively
neglect this contribution according to the discussion in
Reference.~\cite{Bartels:2015uba}.
We find that the IGRB limits with our boost model are competitive to the
dwarf bounds for dark matter massese at $\sim$200~GeV.
Note that more accurate limits should include uncertainties coming from
modeling of the astrophysical contributions.
Further consideration is needed in order to obtain correct values, which
is slated for future works.
(See also Reference.~\cite{Hutten:2017cyu} for a detailed discussion on
various sources of uncertainties.)

The small-scale angular power spectrum of the IGRB has also been
measured with Fermi-LAT~\cite{Fornasa:2016ohl}, which provides yet
another avenue to constrain dark matter annihilation~\cite{Ando:2005xg,
Ando:2013ff} as well as high-energy astrophysical
sources~\cite{Ando:2006cr, Ando:2017alx}.
It is also pointed out that taking cross correlations with local
gravitational tracers such as galaxy catalogs is a promising way along
the same line~\cite{Ando:2013xwa, Ando:2014aoa, Fornengo:2013rga}.
Since these {\it anisotropy} constraints are more sensitive to the dark
matter distribution at smaller redshifts and in larger hosts, the effect
of the subhalo boost is expected to be even more important than for the
IGRB intensity.
A dedicated investigation is beyond the scope of this work and hence
reserved as subject in a future paper.
We also note that our updated boost model will impact the result of
stacking analysis of nearby galaxy groups~\cite{Lisanti:2017qlb}, which
relied on the boost model of Reference.~\cite{Bartels:2015uba}.

\section{Conclustions}
\label{sec:conclusion}

We can access the substructure of dark matter halos which is beyond the
resolutions of the numerical simulations by taking analytical approach
on the modeling of the tidal mass loss of the subhalos. 
We analytically modeled the mass loss of subhalos under the
gravitational potential of their hosts, following the evolution of both
the host and subhalos in a self-consistent way. 
In order to take distributions of the concentrations of the hosts, orbits
and concentrations of subhalos into account, we conducted Monte Carlo
simulations. 
We find that the mass loss of the subhalos are well described with
Eq. ~(\ref{eq:mdot}) down to the scale of $m/M_\mathrm{host}\sim{\cal
O}(10^{-19})$, and well agree with results of $N$-body simulations.

Combining the derived relation about the subhalo mass loss with
analytical models for mass and redshift distributions of accreting
subhalos, we calculated the subhalo mass functions and the boost factor
for dark matter annihilation. 
We showed that mass functions of subhalos derived in our analytical
modeling are consistent with those obtained in $N$-body simulations
down to their resolution limits. 
From our model of the subhalo boost of dark matter annihilation, we
expect enhancement in the gamma-ray signals by up to a factor of
$\sim$10 because of the remaining substructures in larger halos,
predicting promising opportunities for detecting particle dark matter in
future gamma-ray observations.
Including substructures in the subhalos will give important contribution
to the annihiation boost up to a factor of a few.

The results of our calculations are consistent with both earlier
analytical and numerical approaches, but are applicable to much wider
(and arbitrary) range of host masses and redshifts, and hence can be
used to predict gamma-ray flux from dark matter annihilation in various
halos at any redshifts.
As an example, we computed the contribution to the isotropic gamma-ray
background from our boost model.
We find that the presence of subhalos (and their substructures) enhace
the gamma-ray intensity by a factor of a few, and hence the limits on
the annihilation cross section improves by the same factor, excluding
region of $\langle \sigma v\rangle \agt 4\times
10^{-26}$~cm$^3$~s$^{-1}$ for dark matter masses smaller than
$\sim$200~GeV.

\acknowledgments
We thank Richard Bartels for discussions.
This work was supported by JSPS KAKENHI Grant Numbers 17H04836 (SA),
15H01030 and 17H04828 (TI).
Numerical computations were partially carried out on the K computer at
the RIKEN Advanced Institute for Computational Science (Proposal
numbers hp150226, hp160212, hp170231), and Aterui supercomputer at
Center for Computational Astrophysics, CfCA, of National Astronomical
Observatory of Japan.  
TI has been supported by MEXT as ``Priority Issue on Post-K computer''
(Elucidation of the Fundamental Laws and Evolution of the Universe) and
JICFuS.
\appendix 

\section{Mass evolution of host halos}
\label{app:Mass evolution of host halos}

In order to calculate the evolution of subhalos, we first specify how
the hosts that are not in a even larger halo evolve.
Reference.~\cite{Correa:2014xma} derive the relations about the mass
accretion history of the halos $M(z|M_0,z=0)$, i.e., the mass of the
halo at redshift $z$, whose mass is $M_0$ at $z=0$:
\begin{equation}
M(z|M_0,z=0)=M_0(1+z)^\alpha \exp(\beta z),
\label{eq:Mz}
\end{equation}
with
\begin{eqnarray}
\beta&=&-g(M_0), \label{correafbeta}\\
\alpha&=&\left[\frac{1.686\sqrt{2/\pi}}{D^2(z=0)}\left.\frac{dD}{dz}\right|_{z=0}+1\right]g(M_0), \label{correafalpha}\\
g(M_0)&=&\left[S(M_0/q)-S(M_0)\right]^{-1/2}, \label{correaf}\\
q&=&4.137\tilde{z}_f^{-0.9476},\\
\tilde{z}_f&=&-0.0064(\log M_0)^2+0.0237(\log M_0) \nonumber \\
&&+1.8837,
\end{eqnarray}
where $D(z)$ and $S(M)\equiv\sigma^2(M)$ are the growth function and the
variance of the matter distribution at mass scale $M$ and $z = 0$,
respectively.
We adopt fitting functions of both $D(z)$ and $\sigma(M)$ from
Reference.~\cite{Ludlow:2016ifl}.
Eq.~(\ref{eq:Mz}) is generalized to determine the mass of halos
$M(z|M(z_i),z_i)$ at redshift $z$, whose mass was $M(z_i)$ at redshift
$z_i$~\cite{Correa:2015dva}:
\begin{equation}
M(z|M(z_i),z_i)=M(z_i)(1+z-z_i)^\alpha\exp(\beta(z-z_i)) ,
\label{eq:Mzzi}
\end{equation}
with replacing $M_0$ with $M(z_i)$ in Eqs.~(\ref{correafbeta}),(\ref{correafalpha}) and (\ref{correaf}). 
These relations enable us to follow back the evolutions of the hosts
starting from any redshift adopting the generalized equations.

\section{Concentration-mass relation of the field halos}
\label{app:concentration}

We here summarize the concentration-mass relation $c_{200}(M_{200})$ for
the field halos based on Reference.~\cite{Correa:2015dva}, which is adopted
throughout this paper. 
We take the fitted values corresponding to the Planck cosomlogy.

For $z\leq4$
\begin{equation}
\log c_{200}=\alpha +\beta\log\left(\frac{M_{200}}{M_\odot}\right)\left[1+\gamma\log^2\left(\frac{M_{200}}{M_\odot}\right)\right],
\end{equation}
where
\begin{eqnarray}
\alpha&=&1.7543-0.2766(1+z)+0.02039(1+z)^2,\\
\beta&=&0.2753+0.00351(1+z)-0.3038(1+z)^{0.0269},\nonumber\\
 \\
\gamma&=&-0.01537+0.02102(1+z)^{-0.1475},
\end{eqnarray}
and for $z>4$,
\begin{equation}
\log c_{200}=\alpha+\beta\log\left(\frac{M_{200}}{M_\odot}\right),
\end{equation}
where
\begin{eqnarray}
\alpha&=&1.3081-0.1078(1+z)+0.00398(1+z)^2,\\
\beta&=&0.0223-0.0944(1+z)^{-0.3907}.
\end{eqnarray}

\section{Subhalo accretion rate}
\label{ssec:Subhalo accretion rate}

With the understanding of the growth history of certain hosts, we know
the distributions of the mass and redshift of the accreting subhalos on
that host. 
Reference.~\cite{Yang:2011rf} studied the mass accretion history, and
obtained the distribution $d^2N_{\rm sh}/(d \ln m_\mathrm{acc}
dz_\mathrm{acc})$: the number of subhalos accreted onto the host per
unit logarithmic mass range around $\ln m_{\rm acc}$ and per unit
redshift range around accretion redshift $z_{\rm acc}$:
\begin{equation}
\label{maccdistribution}
\frac{d^2N_{sh}}{d\ln m_\mathrm{acc}dz_\mathrm{acc}}={\cal
F}(s_\mathrm{acc},
\delta_\mathrm{acc}|S_0,\delta_0;\overline{M}_\mathrm{acc})\frac{ds_\mathrm{acc}}{dm_\mathrm{acc}}\frac{d\overline{M}_\mathrm{acc}}{dz_\mathrm{acc}},
\end{equation}
where following the convention of Reference.~\cite{Yang:2011rf},
$s_\mathrm{acc}$ and $\delta_\mathrm{acc}$ are used to parameterize the
mass and redshift, respectively, since they are defined as 
$s_\mathrm{acc} \equiv \sigma^2 (m_\mathrm{acc},z=0)$ and
$\delta_\mathrm{acc}=\delta_c(z_\mathrm{acc}) =
1.686/D(z_\mathrm{acc})$~\cite{Ludlow:2016ifl}.
Similarly, for the host, we adopt $S_0 = \sigma^2(M_0,z=0)$ and
$\delta_0 = \delta_{\rm sc}(z_0)$ to characterize the mass $M_0$ and
redshift $z_0$ as a boundary condition.
The mass of the host $M_{\rm acc}$ at the accretion redshift $z_{\rm
acc}$ (that eventually evolves to $M_0$ at $z_0$) follows the
probability distribution $P(M_{\rm acc}| S_0, \delta_0)$, for which we
adopt a log-normal distribution with a logarithmic mean
$\overline{M}_\mathrm{acc}=M(z_\mathrm{acc}|M_0,z_0)$
[Eq.~\ref{eq:Mzzi}] and a logarithmic dispersion
\begin{equation}
\sigma_{\log
 M_{\rm acc}}=0.12-0.15\log\left(\frac{{M}_\mathrm{acc}}{M_0}\right).
\end{equation}
The definition of the function ${\cal F}$ in
Eq.~(\ref{maccdistribution}) is 
\begin{eqnarray}
&&{\cal
 F}(s_\mathrm{acc},\delta_\mathrm{acc}|S_0,\delta_0;\overline{M}_\mathrm{acc})
 \nonumber \\
&=&\int \Phi (s_\mathrm{acc}.\delta_\mathrm{acc}|S_0,\delta_0;{M}_\mathrm{acc})P({M}_\mathrm{acc}|S_0,\delta_0)d{M}_\mathrm{acc},\nonumber\\
\end{eqnarray}
\begin{eqnarray}
&&\Phi(s_\mathrm{acc},\delta_\mathrm{acc}|S_0,\delta_0;{M}_\mathrm{acc}) \nonumber \\
&=&\left[\int^\infty_{S(m_\mathrm{max})}F(s_\mathrm{acc},\delta_\mathrm{acc}|S_0,\delta_0;{M}_\mathrm{acc})ds_\mathrm{acc}\right]^{-1} \nonumber \\
&&\times\begin{cases} F(s_\mathrm{acc},\delta_\mathrm{acc}|S_0,\delta_0;{M}_\mathrm{acc}),& (m_\mathrm{acc}\leq m_\mathrm{max}), \\
0, & (\mathrm{otherwise}), 
\end{cases}\nonumber\\
\end{eqnarray}
\begin{eqnarray}
&&F(s_\mathrm{acc},\delta_\mathrm{acc}|S_0,\delta_0;{M}_\mathrm{acc}) \nonumber \\
&=&\frac{1}{\sqrt{2\pi}}\frac{\delta_\mathrm{acc}-\delta_M}{(s_\mathrm{acc}-S_M)^{3/2}}\exp\left[-\frac{(\delta_\mathrm{acc}-\delta_M)^2}{2(s_\mathrm{acc}-S_M)}\right],
\end{eqnarray}
where $m_\mathrm{max}=\min[{M}_\mathrm{acc},M_0/2]$ and
$M_\mathrm{max}=\min[{M}_\mathrm{acc}+m_\mathrm{max},M_0]$ are
introduced such that the mass hierarchy of the host mass before and
after subhalo accretions is assured, $S_M=\sigma_M^2(M_\mathrm{max})$
and $\delta_M$ is defined as $\delta_{\rm sc}(z)$ at a redshift at
which $M = M_{\rm max}$.
The equations above determine the distributions of accreting subhalos
$d^2N_\mathrm{sh}/(d\ln m_\mathrm{acc}dz_\mathrm{acc})$ for arbitrary
hosts.

\bibliographystyle{utphys}
\bibliography{refs}

\end{document}